\documentclass[prd,amsmath,amssymb,floatfix,superscriptaddress,nofootinbib,twocolumn]{revtex4-1}

\newcommand{\vn}{\vec{n}}
\newcommand{\tdX}{\tilde X}
\newcommand{\bea}{\begin{eqnarray}}
\newcommand{\eea}{\end{eqnarray}}
\newcommand{\vl}{\vec{\ell}}
\newcommand{\ec}[1]{Eq.~(\ref{eq:#1})}

\newcommand{\eql}[1]{\label{eq:#1}}
\newcommand{\vL}{\vec{L}}

\usepackage{calligra}
\DeclareMathAlphabet{\mathcalligra}{T1}{calligra}{m}{n}
\DeclareMathAlphabet{\mathpzc}{OT1}{pzc}{m}{it}
\usepackage[utf8]{inputenc}
\usepackage{graphicx}
\usepackage{amssymb}
\usepackage{simplewick}
\usepackage{amsmath}
\usepackage{bm}
\usepackage{color}
\usepackage{enumitem}
\usepackage[linktocpage=true]{hyperref} 
\hypersetup{
    colorlinks=true,       
    linkcolor=red,         
    citecolor=blue,        
    filecolor=magenta,      
    urlcolor=blue           
}
\usepackage[all]{hypcap} 
\definecolor{darkgreen}{cmyk}{0.85,0.1,1.00,0} 
\definecolor{darkorange}{rgb}{1.0,0.2,0.0}

\begin{document}

\title{High-Resolution CMB Lensing Reconstruction with Deep Learning}

\author{Peikai Li}
\email{peikail@andrew.cmu.edu}
 \affiliation{Department of Physics, Carnegie Mellon University, Pittsburgh, PA 15213, USA}
 \affiliation{McWilliams Center for Cosmology, Carnegie Mellon University, Pittsburgh, PA 15213, USA}
 \affiliation{NSF AI Planning Institute, Carnegie Mellon University, Pittsburgh, PA 15213, USA}

\author{Ipek Ilayda Onur}%
 \email{ionur@andrew.cmu.edu}
\affiliation{Department of Electrical and Computer Engineering, Carnegie Mellon University, Pittsburgh, PA 15213, USA}%

\author{Scott Dodelson}
\email{sdodelso@andrew.cmu.edu}
 \affiliation{Department of Physics, Carnegie Mellon University, Pittsburgh, PA 15213, USA}
 \affiliation{McWilliams Center for Cosmology, Carnegie Mellon University, Pittsburgh, PA 15213, USA}
 \affiliation{NSF AI Planning Institute, Carnegie Mellon University, Pittsburgh, PA 15213, USA}

\author{Shreyas Chaudhari}
 \email{shreyasc@andrew.cmu.edu}
\affiliation{Department of Electrical and Computer Engineering, Carnegie Mellon University, Pittsburgh, PA 15213, USA}%

\date{\today}

\begin{abstract}
Next-generation cosmic microwave background (CMB) surveys are expected to provide valuable information about the primordial universe by creating maps of the mass along the line of sight. Traditional tools for creating these lensing convergence maps include the quadratic estimator and the maximum likelihood  based iterative estimator. Here, we apply a generative adversarial network (GAN) to reconstruct the lensing convergence field. We compare our results with a previous deep learning approach -- Residual-UNet -- and discuss the pros and cons of each. In the process, we use training sets generated by a variety of power spectra, rather than the one used in testing the methods.
\end{abstract}

\maketitle

\section{\label{sec:level1}Introduction}
By observing the cosmic microwave background (CMB) we learn about the primordial universe \cite{Hu:2008hd}. The CMB photons are deflected by the intervening gravitational potential along the line of sight between the last scattering surface and the Earth \cite{Lewis:2006fu}. This phenomena is known as CMB gravitational lensing and we can extract important information about the gravitational potential along with the properties of the universe from it \cite{Planck:2018lbu}. Hu pointed out that the projected potential $\phi$ can be reconstructed using a quadratic estimtor (QE) from the observed CMB temperature and polarization fields \cite{Hu:2001tn}. The QE is almost optimal at high noise levels. However, current and upcoming CMB surveys, e.g. \cite{Henderson:2015nzj}\cite{SPT-3G:2014dbx}\cite{SimonsObservatory:2018koc}\cite{Hui:2018cvg}\cite{Aravena:2019tye}\cite{CMB-S4:2016ple}\cite{NASAPICO:2019thw}, are designed to be capable of reaching unprecedented low levels of noise. These surveys will provide us with a significantly higher CMB lensing reconstruction signal-to-noise ratio \cite{Maniyar:2021msb}, which in turn will better constrain the cosmological parameters. Yet the QE is found to be sub-optimal for these surveys with low noise level ($\sim$ a few $\mu$K-arcmin) \cite{Hirata:2003ka}. Therefore the community has been actively developing analytical approaches to tackle this problem, e.g. the iterative estimator \cite{2012JCAP...06..014S}, the gradient-inversion method \cite{Hadzhiyska:2019cle}, likelihood-based methods \cite{Hirata:2003ka}\cite{Carron:2017mqf}\cite{Millea:2017fyd} and recently the global-minimum-variance estimator \cite{Maniyar:2021msb}. Almost all these methods are built upon the vanilla QE, which, despite being sub-optimal, will still play an important role in recent surveys since it is easy and fast to implement computationally.

In our work, we demonstrate that deep learning can be a powerful tool for lensing reconstruction. Recently, there have been papers \cite{Caldeira:2018ojb}\cite{Guzman:2021nfk} trying to apply the Residual-UNet (ResUNet) \cite{2017arXiv170103056K}\cite{2018IGRSL..15..749Z} to the CMB convergence $\kappa$ map reconstruction task\footnote{The convergence $\kappa$ map only differs from the projected gravitational potential field $\phi$ by a constant field in Fourier space, so essentially the task is still reconstructing the lensing potential.}. The ResUNet model is based on convolutional neural network (CNN) and is one of the most reliable architecture in the field of computer vision. ResUNet achieved great success in reconstructing the primordial $\tilde B$ field and the lensing convergence $\kappa$ field, further it was proved to be capable of providing a lower lensing reconstruction noise than the QE approach when the detector noise is low. However ResUNet struggled to reconstruct the convergence field to a high precision in the presence of non-negligible noise. We realize that the reconstructed convergence $\hat\kappa$ field under such non-negligible detector noise can be viewed as a low-resolution reconstruction of the true convergence map. Therefore in this work, we seek help from the super-resolution image reconstruction architecture $Pix2PixHD$ model \cite{2017arXiv171111585W} in deep learning. $Pix2PixHD$ model is a powerful tool of reconstructing high-resolution images from low-resolution ones based on generative adversarial network (GAN) \cite{2014arXiv1411.1784M}. We develop a $Pix2PixHD$ based deep learning architecture that is capable of predicting the CMB convergence map $\kappa$ to a high-resolution, and the the reconstructed lensing power spectrum is almost perfectly aligned with the ground truth power spectrum despite detector noise. We make two major modifications to the $Pix2PixHD$ model: first, we use the observed CMB Stokes parameters $Q$ and $U$ as the input, where as in most super-resolution networks the input is the low-resolution version of the target image; second we get rid of the conditional assumption and also add Fourier space loss in the discriminator loss. We will talk about these in detail later. We also compare the performance of the ResUNet and our GAN model using a generalized dataset, where we vary the analytical convergence power spectrum $C_L^{\kappa\kappa}$ when constructing the set of $\{((Q, U), \kappa) \}$ realizations. By doing this, we wish to test the robustness of these models: how will they extract an unknown power spectrum.

We organize the paper as follows: in section \ref{sec:level2} we briefly introduce the background of CMB lensing reconstruction. In section \ref{sec:level4} we introduce our GAN model: the modified version of $Pix2PixHD$ network. In section \ref{sec:level6} we introduce the dataset we use and compare the results from the ResUNet and the GAN model using this dataset. Finally we conclude the work in section \ref{sec:level7}. The source code of the work can be found in \url{https://github.com/ionur/CMB}.

\section{\label{sec:level2}CMB Lensing}
In this section we briefly review the theory of CMB lensing reconstruction: the analytical background of the reconstruction, the input and target maps of the neural net and the detector noise model we apply. We will be using the flat-sky approximation throughout this work, also we use $\vec \ell$ to denote Fourier wavenumber for CMB fields and $\vec L$ for the lensing potential, as in \cite{Maniyar:2021msb}.

The observed CMB field $X(\vn)$ in direction $\vn$ differs from the primordial CMB field $\tdX(\vn)$ by a deflection angle $\vec d$. Here $ X=T, E, B$ can be either the temperature or the polarization field. This deflection angle $\vec d$ is related to the projected gravitational potential $\phi$ as $\vec d(\vn)=\nabla \phi(\vn)$. The observed CMB field $X$ is no longer isotropic\footnote{We ignore primordial non-Gaussianity throughout this work.}, due to the existence of the deflection field. Analytically the result of this is that non-zero off-diagonal elements of the following two-point functions will no longer vanish:
\bea 
\langle X(\vl_1)Y(\vl_2)\rangle &=& (2\pi)^2\delta_{2\rm D}(\vl_1+\vl_2)C_{\ell_1}^{XY}\nonumber \\
& & + \textrm{off-diagonal elements}, 
\eea 
here $X, Y=T, E, B$ are the observed CMB fields in Fourier space, and $C_{\ell}^{XY}$ is the diagonal element of the covariance, which is the power spectrum of the CMB. The non-zero off-diagonal elements carry essential information about the deflection field $\vec d$, which is directly related to the lensing potential $\phi$. Thus we can express the $\phi$ field in Fourier space as quadratic pairs of $X$ and $Y$. The convergence field $\kappa$ is related to $\phi$ in Fourier space by:
\bea 
\kappa(\vec L) = -\frac{L^2}{2} \phi(\vec L)
\eea 
so the quadratic estimator can also be applied to the convergence field:
\bea 
\hat \kappa^{\rm QE}_{XY}(\vL)=\sum_{\vl_1+\vl_2=\vL}X(\vl_1)Y(\vl_2)F_{XY}(\vl_1,\vl_2) \eql{QE}.
\eea 
\ec{QE} is the equation for the quadratic estimator where $X$ and $Y$ are Fourier modes of observed $T, E, B$ maps, $F_{XY}$ can be fully determined analytically using both observed and primordial (auto or cross) power spectrum of the temperature and polarization fields. This expression \ec{QE} is the analytical function mapping the input maps to the target $\kappa$ map.

The observed CMB maps contain various types of noise, including detector noise, which will be the main concern of this work. Detector noise is taken to be Gaussian and homogeneous, with power spectra given by \cite{PhysRevD.52.4307}:
\bea 
C_{\ell}^{TT}|_{\rm noise}&=&\Delta_T^2e^{\ell(\ell+1) \theta_{\rm FWHM}/8\ln 2 }, \\
C_{\ell}^{EE}|_{\rm noise}&=&C_{\ell}^{BB}|_{\rm noise}=\Delta_P^2e^{\ell(\ell+1) \theta_{\rm FWHM}/8\ln 2 },
\eea 
where $\Delta_T$ and $\Delta_P$ determine detector noise, and $\theta_{\rm FWHM}$ is the FWHM of the beam.

Similar to previous work \cite{Caldeira:2018ojb}\cite{Guzman:2021nfk}, we use the observed Stokes parameters $Q$ and $U$ as the input. The $Q$ and $U$ maps are related to the polarization $E$ and $B$ maps in Fourier space by \cite{Zaldarriaga:2001st}:
\bea 
Q(\vl)&=&E(\vl)\cos(2\phi_\ell)-B(\vl)\sin(2\phi_{\ell})\\
U(\vl)&=&E(\vl)\sin(2\phi_\ell)+B(\vl)\cos(2\phi_{\ell})
\eea 
where $\phi_\ell=\cos^{-1}(\hat{{x}}\cdot \hat{{\ell}})$. Thus $Q$ and $U$ maps encode full information of observed CMB polarization maps. The target map is the CMB convergence map $\kappa$.

\section{\label{sec:level4}Modified $Pix2PixHD$ Model}

In this section we introduce the details of the deep learning architecture we use in this work: the modified $Pix2PixHD$.

The $Pix2PixHD$ model is one of the most promising super-resolution deep learning methods. This model is based on conditional generative adversarial network (cGAN), which is composed of two parts: a generative network (generator $G$) and a  discriminative network (discriminator $D$). The generator has a CNN-based structure and is aiming at reconstructing the super-resolution image. Multiple discriminators try to distinguish the generated image from the ground truth image at multiple scales, considering different patches.

\begin{figure*}[ht!]
    \centering
    \includegraphics[width=\textwidth]{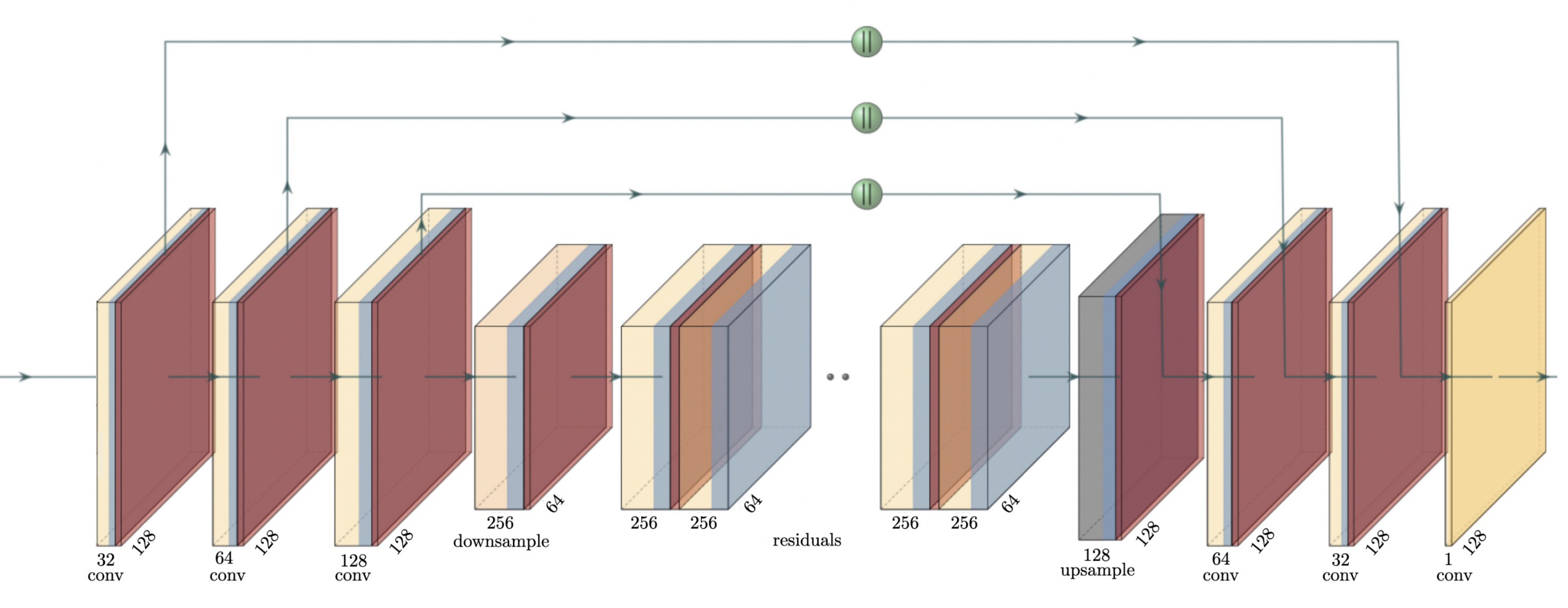}
    \includegraphics[width=0.45\textwidth]{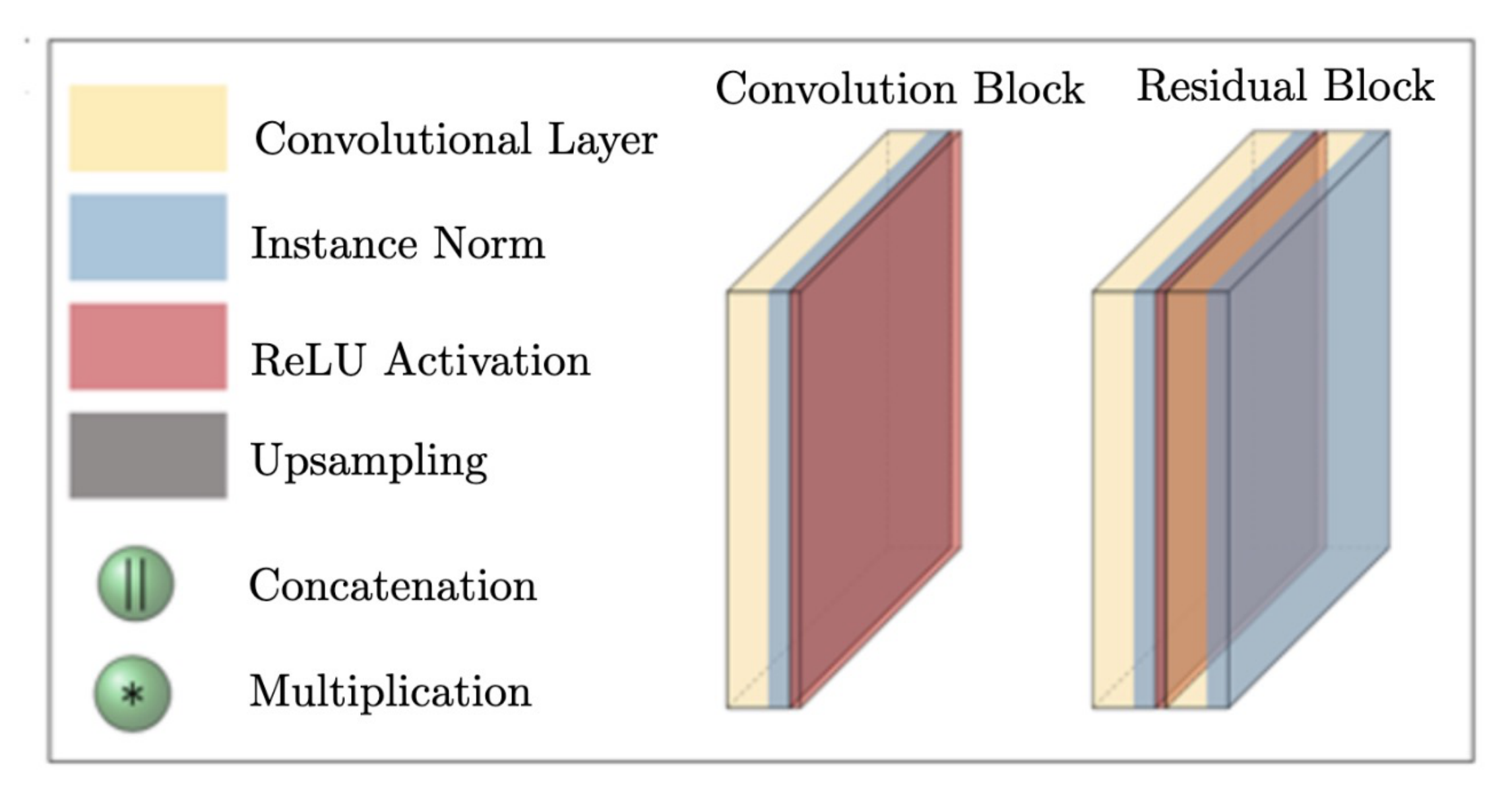}
    \caption{Modified $Pix2PixHD$ model's generator architecture. 
    The input is the concatenated ($Q$, $U$) CMB maps and 
    $\hat{\kappa}$ map is the output of the generator. We multiply the output of the last hidden layer by the apodization mask to get the final result. The number of filters of each layer is shown underneath each block as well as the image size. 
    Each convolution block has instance normalization followed by the rectified linear unit (ReLU) activation except for the last layer. 
    There are 5 residual blocks forming the local enhancer $G_2$.
    (The graphic is created from the publicly available \url{https://github.com/HarisIqbal88/PlotNeuralNet} repository.)}
    \label{fig:generator}
\end{figure*}

The training set for the CMB lensing reconstruction task is a set of images $\{ (S_i,\kappa_i)  \}_{i=1}^N$, where $N$ is the number of training samples and $i$ is the index for a specific realization. The input $S_i = (Q_i,U_i)$ can be considered as an image with two channels, $Q_i$ and $U_i$. The target map $\kappa_i$ is the high-resolution image we want to reconstruct. The loss function of GAN is typically given by:
\bea 
\mathcal{L}_{\rm GAN}=\mathbb{E}_{\kappa}[\log D(\kappa)]+ \mathbb{E}_{S}[\log(1-D(G(S)))].
\eea  
where $\mathbb{E}_{\kappa}$ and $\mathbb{E}_S$ are the expectation value over all target maps and all input maps, respectively. The output of the generator $G(S)$ is the predicted convergence map $\hat \kappa$. The generator $G$ tries to minimize the loss function while the discriminator tries to maximize it:
\bea 
\min_G \max_D \mathcal{L}_{\rm GAN}(G,D).
\eea 
Notice we remove the conditional assumption, since unlike usual super-resolution networks, here the input is the Stokes parameters $Q$ and $U$ instead of the low-resolution version of the target $\kappa$. Therefore the conditional assumption will actually worsen the performance of the discriminator.

Like the $Pix2PixHD$ model, we decompose the generator into two sub-networks: $G_1$ and $G_2$. $G_1$ is considered as the global generator network and $G_2$ as the local enhancer network. The global generator consists of two blocks: one downsampling block and one upsampling block. For the downsampling block, we use two convolutional layers with stride $1$ followed by a downsampling layer with stride $2$. Correspondingly, the upsampling layer is a transposed convolutional upsampling layer with stride $2$ followed again by two convolutional layers with stride $1$. Since the goal is to reconstruct a high resolution $\hat \kappa$ map and our map size is relatively small ($128\times 128$ pixels), we found that the stride $2$ downsampling-upsampling combination more than once would make the output map worse. The local enhancer network we have is a set of ($5$) residual blocks. We also concatenate the output from the upsampling block with those from the downsampling block, which is the skip connection. The detailed structure of the GAN's generator is shown in Figure~\ref{fig:generator}.

\begin{figure*}[ht!]
    \centering
    \includegraphics[width=\textwidth]{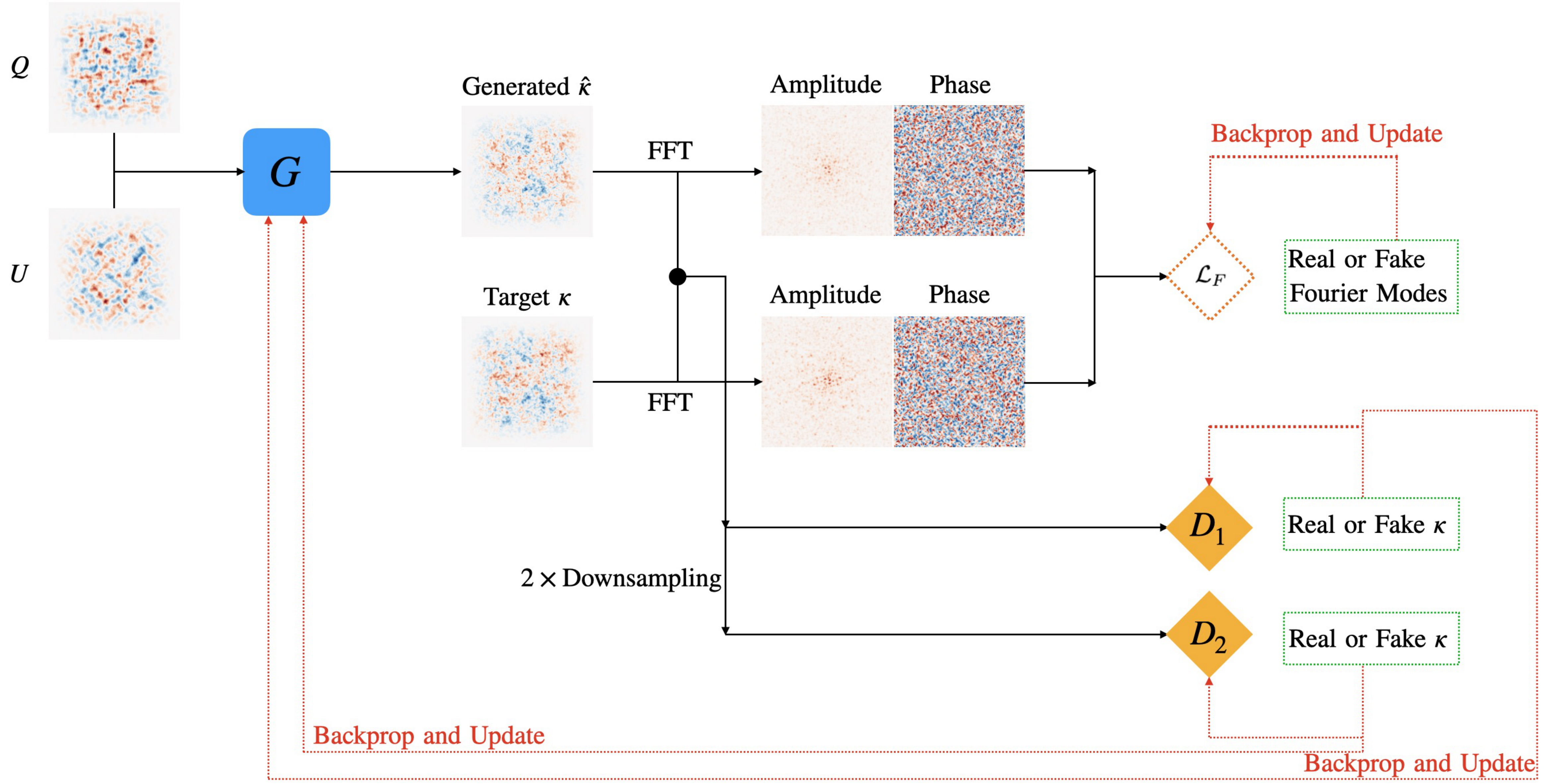}
    \caption{Flowchart and structures of the modified $Pix2PixHD$ model.Here $G$ is the generator, $D_1$ and $D_2$ are the discriminators. Here the Fourier space amplitude map is $|\hat \kappa(\vec L)|^2L$ (or $| \kappa(\vec L)|^2L$).}
    \label{fig:modifiedPix2pixHD}
\end{figure*}

We use a discriminator structure similar to that in \cite{2020ApJ...895L..16S}: two discriminators $D_1, D_2$ with identical architecture. The discriminator $D_1$ gets input images with the original pixel size, while the other one $D_2$ gets downsampled by half input images. These two discriminators focus on the differences of the generated $\hat \kappa$ and the true $\kappa$ maps at different scales. The $D_1$ discriminator has a global view of the image, making sure that the generated $\hat \kappa$ has correct global properties, while $D_2$ pays more attention to the smaller scale consistency. We also add a feature matching loss to the GAN loss in order to improve the generator training:
\bea 
\mathcal{L}_{\rm FM}(G,D_i)=\mathbb{E}_{(S,\kappa)}\sum_{l=1}^{L_D}\frac{1}{N_l}[|| D_i^{(l)}(\kappa)-D_i^{(l)}(G(S)) ||_1] \nonumber \\
\eea 
where $l$ represent the $l$-th layer of the discriminator $D_i$ and $L_D$ is the total number of layers in each discriminator. Now the full minimax algorithm becomes:
\bea 
\min_G \bigg[ \bigg( \max_{D_1,D_2}\sum_{i=1,2}\mathcal{L}_{\rm GAN}(G, D_i) \bigg) +\lambda \sum_{i=1,2}\mathcal{L}_{\rm FM}(G, D_i) \bigg]. \nonumber \\
\eea 
We use $10$ for a relative weight $\lambda$ which determines the importance of $\mathcal{L}_{\rm GAN}$ and $\mathcal{L}_{\rm FM}$ as in \cite{2017arXiv171111585W}. Notice that the feature matching loss will only be effective when we back-propagate through the generator. 

As a further improvement to the discriminator network, we have introduced another discriminator loss $\mathcal{L}_F$ that computes the Fourier space mean-squared-error (MSE) loss between the predicted $\hat \kappa$ and the true $\kappa$ maps. We take the Fast Fourier Transform of both the $\hat \kappa$ and the true $\kappa$ maps and obtain an amplitude map and a phase map $\angle$ for each. We multiply the amplitude map by $L$ to focus more on high $L$ differences. The Fourier space loss is given by
\bea
&&\mathcal{L}_{F}(G)=\mathbb{E}_{(S,\kappa)}\bigg[\lambda_{\rm amp}|| |\kappa(\vec L)|^2L-|\hat \kappa(\vec L)|^2L ||_2 \nonumber\\
&&\qquad \qquad +\lambda_{\rm phase}|| \angle(\kappa(\vec L))-\angle(\hat \kappa(\vec L)) ||_2 \bigg]\nonumber\\ 
\eea
This loss depends on $G$ since $\hat \kappa = G(S)$. In practice, we find that including the Fourier space loss in the discriminator loss is better than directly adding it to the generator loss. The relative weights are set to be $\lambda_{\rm amp}=20$ and $\lambda_{\rm phase}=1/20$ respectively, to maximize the performance of the network. Now the full objective function becomes:
\bea
&&\min_G  \bigg( \max_{D_1,D_2}\sum_{i=1,2}\mathcal{L}_{\rm GAN}(G, D_i)+ \ \mathcal{L}_{F}(G) \bigg)\nonumber\\
&+&\lambda \sum_{i=1,2}\mathcal{L}_{\rm FM}(G, D_i) 
\eea
The modified Pix2PixHD model's architecture is illustrated in Figure~\ref{fig:modifiedPix2pixHD}.

\section{\label{sec:level6}Data and Results}
\subsection{Data}
We generate simulated data to build and train the deep learning models. We mostly follow the data pipeline in \cite{Guzman:2021nfk} for our data preparation \footnote{\url{https://github.com/EEmGuzman/resunet-cmb}}. The cosmological parameters we use in this work are $H_0=67.9\rm km\, s^{-1}\textrm{Mpc}^{-1}$, $\Omega_b\, h^2=0.0222$, $\Omega_c \,h^2=0.118$, $n_s=0.962$, $\tau=0.0943$ and $A_s=2.21\times 10^{-9}$, in order to compare with the results of \cite{Caldeira:2018ojb}\cite{Guzman:2021nfk}. The analytical convergence power spectrum $C_L^{\kappa\kappa}$ mentioned below is computed using this set of parameters. This data pipeline first uses \textbf{CAMB}\footnote{\url{https://camb.info}} to generate the theoretical power spectra, then uses a modified version of \textbf{Orphics} \footnote{\url{https://github.com/msyriac/orphics}} to generate CMB maps for training. 

Previous work \cite{Caldeira:2018ojb}\cite{Guzman:2021nfk} used a fixed lensing power spectrum to generate observed CMB polarization maps. In this work, in order to further test the robustness of the models, we generate the observed $Q$ and $U$ maps using $\alpha C_{\ell}^{\kappa \kappa}$ where $\alpha$ is a multiplication factor randomly generated within range $0.75-1.25$. We add the simplest form of uncertainty to the target $\kappa$ maps, since in real surveys we will not know the actual lensing power spectrum. In future work, we will vary the power spectrum in a more realistic manner (e.g., by varying the baryon density) and train the network with a more complex dataset.

Each generated map has a size of $5^\circ\times 5^{\circ}$ and $128\times 128$ pixels, with a cosine taper of $1.5^\circ$ to get rid of the effect of periodic boundary conditions in Fourier space. We choose $4$ noise levels with $\Delta_T=0.0 \, \mu K$-$\textrm{arcmin}$, $\Delta_T=1.0 \, \mu K$-$\textrm{arcmin}$, $\Delta_T=2.0 \, \mu K$-$\textrm{arcmin}$ and $\Delta_T=5.0 \, \mu K$-$\textrm{arcmin}$ with $\Delta_P =\sqrt{2} \Delta_T$ and a beam smoothing with size $1$-arcmin. (Notice \cite{Guzman:2021nfk} uses a beam size of $1.4$-arcmin, however due to the size and resolution of our maps, this difference is negligible.) We generate $40,000$ samples of `observed' $Q$, $U$ and ground truth $\kappa$ maps for each noise level. We include the patchy reionization effect \cite{Hu:1999vq} in our dataset although we do not plan to reconstruct it in this work. We train the ResUNet and our model seperately with these $40,000$ samples and compare the result. The train-validation-test split ratio is set to be $90:5:5$.
\begin{widetext}
   \begin{figure*}
     \centering
        \includegraphics[width=15cm,height=10cm]{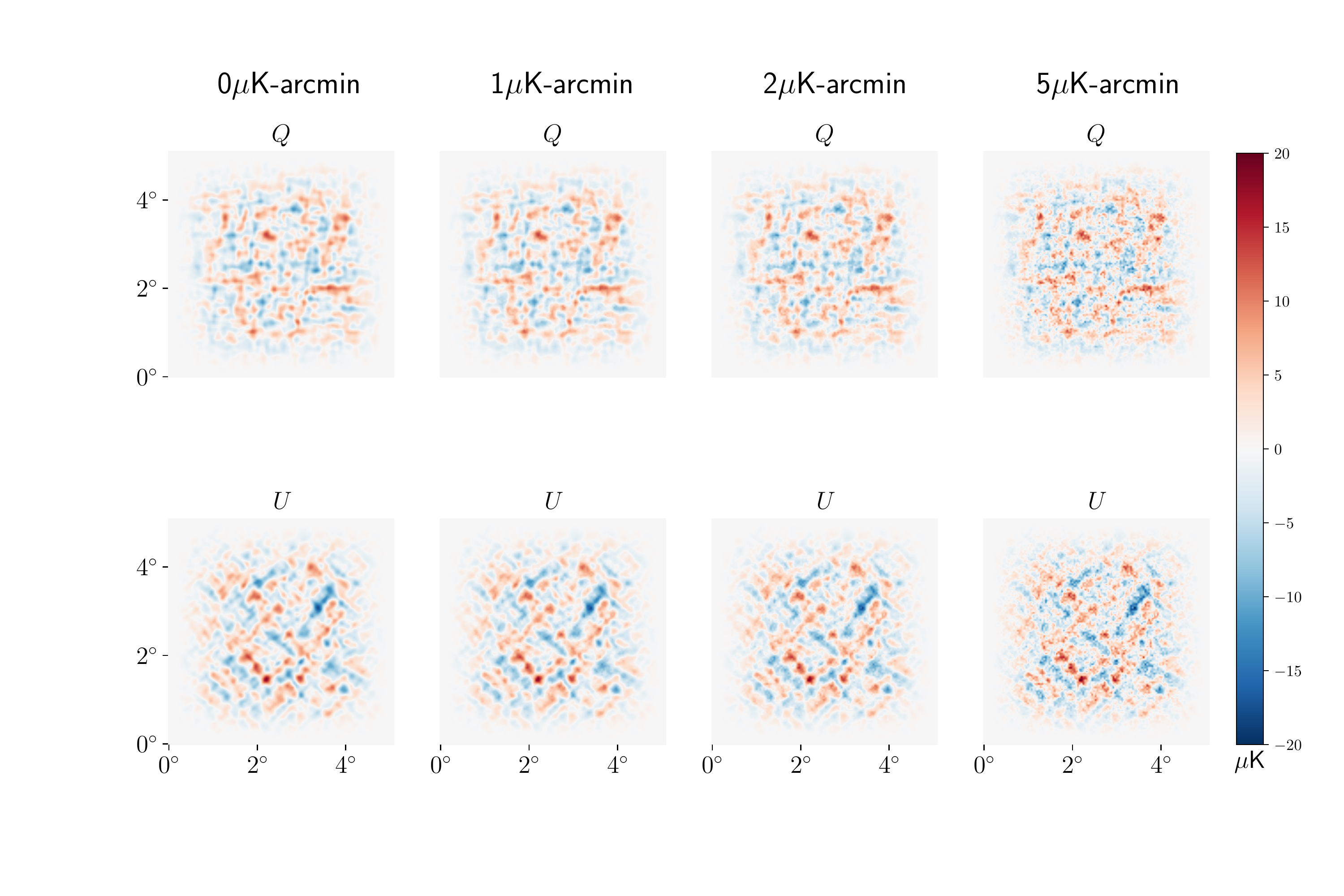}
    \caption{Example maps of the observed CMB stokes parameters $Q$ (first row) and $U$ (second row) with different amounts of noise. Notice that all maps are multiplied by a window function -- the apodization mask, in order to remove the unphysical periodic boundary conditions during the map generating process. All $Q$ (or $U$) maps in this figure are generated with a fixed random seed, so the only difference between them is the detector noise.}
    \label{fig:eg_qu_maps}
   \end{figure*}
\end{widetext}
   
We show example input maps ($Q$ and $U$) with different noise levels in Figure~\ref{fig:eg_qu_maps}. The target map, which is the corresponding true $\kappa$ field can be found in the top-left corner of Figure~\ref{fig:eg_kappa_maps}. Except for the highest noise level we consider ($5\mu$K-armin), it seems that the effect of the detector noise appears to be negligible in Figure~\ref{fig:eg_qu_maps}; i.e., the first $3$ maps in each column appear identical. However in the next subsection \ref{sec:level6p2} we will see that even the smallest amount of detector noise makes the reconstruction nontrivial.

\clearpage

\subsection{\label{sec:level6p2}Training Results: Predicted Maps}
\begin{widetext}
   \begin{figure*}
     \centering
        \includegraphics[width=18cm,height=18cm]{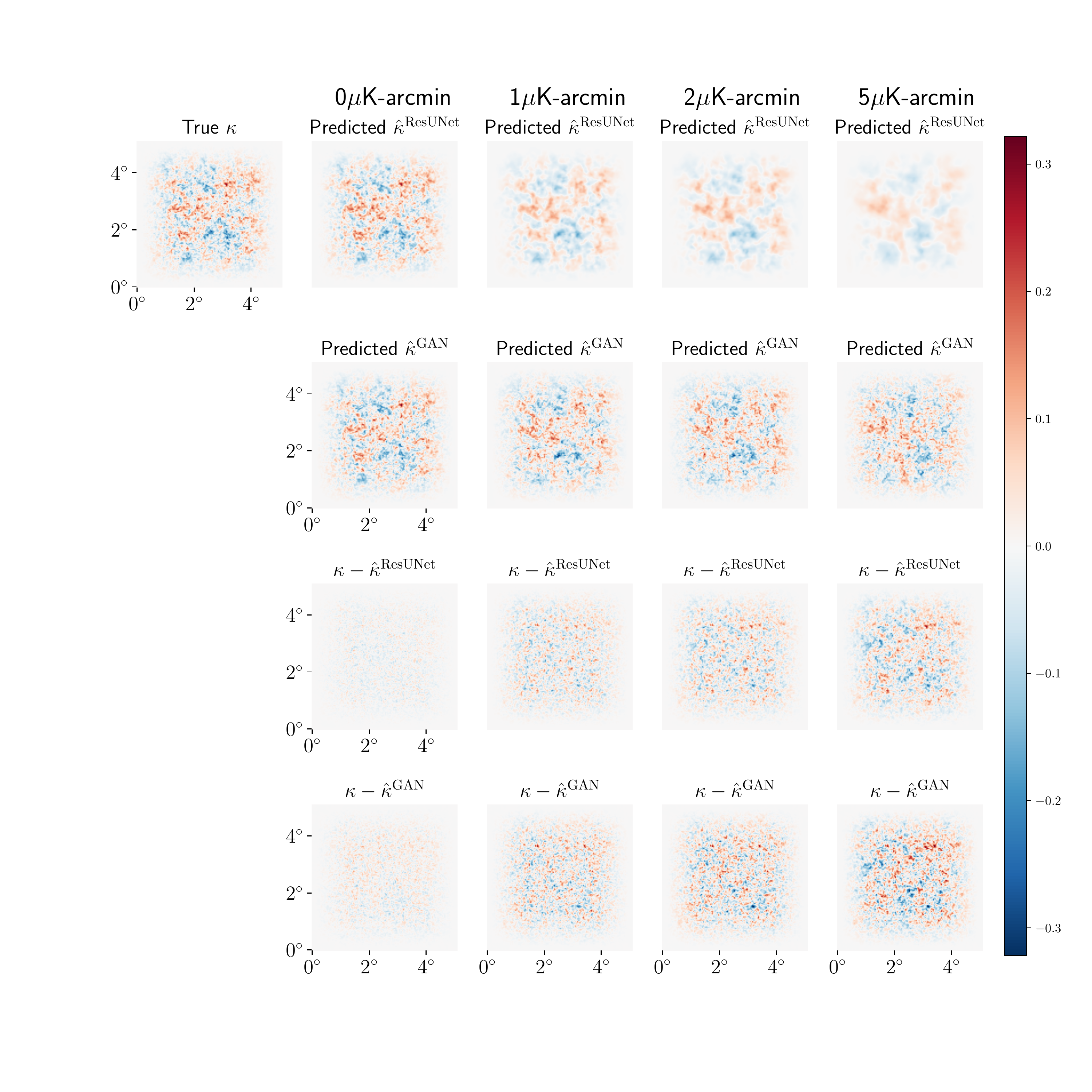}
    \caption{The reconstructed $\hat \kappa$ maps versus the true target $\kappa$ map. The top-left map is the ground truth map, and the remaining $4$ maps in the first row are the reconstructed $\kappa$ maps using the ResUNet model. The $4$ maps in the second row are the GAN model output, and the last two rows are the difference maps between the true field and the two types of reconstructed fields. Each of the $4$ column to the right corresponds to a specific detector noise level.}
    \label{fig:eg_kappa_maps}
   \end{figure*}
\end{widetext}


In this work we use Google Colab Pro $+$ to train our models \footnote{\url{https://colab.research.google.com}}. To further assess the quality of the generated maps, we train the two models -- ResUNet and the GAN model separately for each noise level. For the ResUNet training, we use the same training pipeline and the same set of hyperparameters as in \cite{Guzman:2021nfk}. The batch size is $32$ and the initial learning rate is $0.25$, with a $0.5$ decay rate if there is no improvement in validation loss for three successive epochs. The training will stop when there is no improvement in validation loss for ten consecutive epochs and the model will be saved at the lowest validation loss.

For the training of our GAN model, we also use the Adam optimizer \cite{2014arXiv1412.6980K} and we set the hyperparameters as follows: the batch size is $32$, $\beta_1=0.5$, and the learning rate is set to be $0.0003$. Here $\beta_1$ is the initial decay rate for estimating the first moment of the gradient when using the Adam optimizer. For the GAN training, we track the performance of the model training every $10$ steps by computing and recording some important features (e.g., the mean-squared-error loss of the binned power spectrum $L_{2}^{\rm PS}$, and the signal-to-noise ratio $\big( \frac{S}{N} \big)^2_{L}$ in Fourier space) on the validation set. We use the model if its validation output provides us with both a low $L_2^{\rm PS}$ and a high $\big( \frac{S}{N} \big)^2_{L}$.

   \begin{figure}
     \centering
        \includegraphics[width=9cm,height=12cm]{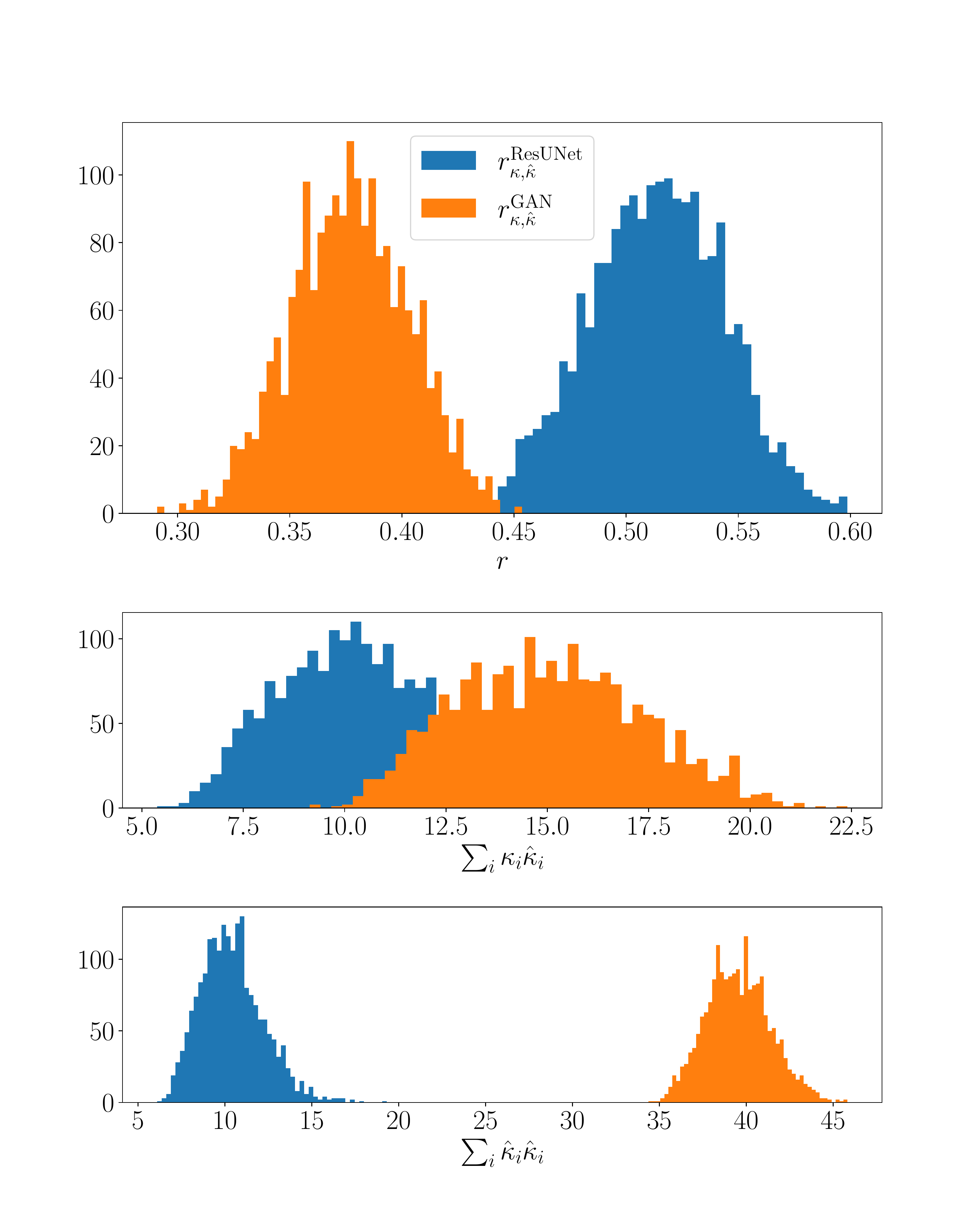}
    \caption{First row is the histogram of the correlation between predicted maps and the true maps for the test set, for both the ResUNet and the GAN model. Second row is the histogram of $\sum_i \kappa_i \hat\kappa_i$ and the last row is the histogram of $\sum_i \hat\kappa_i \hat\kappa_i$ for all realizations in the test set. Noise level of the figure is taken to be $2\mu$K-arcmin. Blue and orange correspond to the ResUNet and the GAN, respectively.}
    \label{fig:r}
   \end{figure}
   
We first compare the result of the ResUNet and our GAN model directly at the map level in Figure~\ref{fig:eg_kappa_maps}. We can see that the ResUNet struggles to reproduce high resolution convergence maps in the presence of detector noise. In those cases, small scale features are missing in the $\hat \kappa^{\rm ResUNet}$ maps. The GAN model does recover some of these small scale features of the map while not losing track of the global properties. For the difference maps however, the difference on the fourth row (GAN) is slightly greater than the difference on the third row (ResUNet). This is actually predictable since ResUNet uses the MSE loss as the optimization objective (which is the sum of the squared value of the pixels in the difference map), while this MSE error is not the only goal of the minimax optimization of the GAN model. The GAN model sacrifices the optimality in MSE loss in order to capture small scale features. We will see more evidence of this trade-off shortly.

Another way to describe the quality of the reconstructed maps is the correlation. The correlation between two maps in real space is defined as:
\bea 
r_{\kappa, \hat\kappa} = \frac{\sum_i \kappa_i \hat\kappa_i}{\sqrt{(\sum_i \kappa_i)^2(\sum_i\hat \kappa_i)^2}},
\eea 
where the summation is performed over all pixels $i$ of the maps. In Figure~\ref{fig:r} we show the histogram of the correlation $r$ for both models on the top row, the detector noise is taken to be $2\mu$K-arcmin (the performance is similar for other noise levels). We see that the correlation for the GAN model is lower than the ResUNet model. However notice that since the GAN model is putting back all the small scale power of the spectrum, the denominator of $r_{\kappa, \hat\kappa}^{\rm GAN}$ is much larger than that of $r_{\kappa, \hat\kappa}^{\rm ResUNet}$, as shown on the third row of Figure~\ref{fig:r}. Besides the numerator of the correlation for GAN is actually slightly larger than that of ResUNet, see the second row of Figure~\ref{fig:r}. We can understand this as: during putting back the small scale features when training the GAN model, we are actually putting back `signal' and `noise' simultaneously. In the next few subsections, we will further evaluate and compare the quality of the predicted maps.
   
\subsection{Training Results: Reconstructed Power Spectrum}
\begin{figure}
     \centering
        \includegraphics[width=9cm,height=6cm]{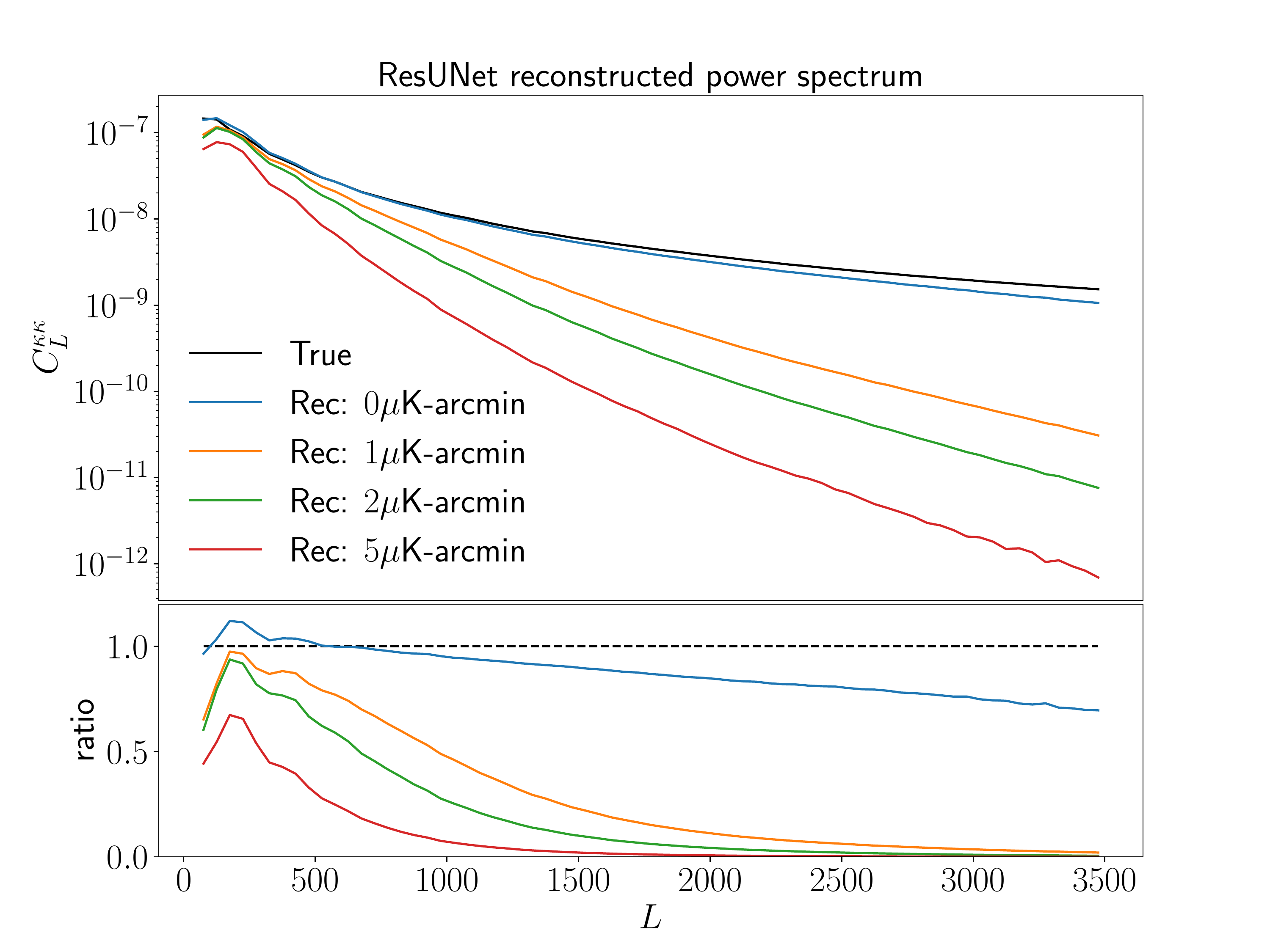}
        \includegraphics[width=9cm,height=6cm]{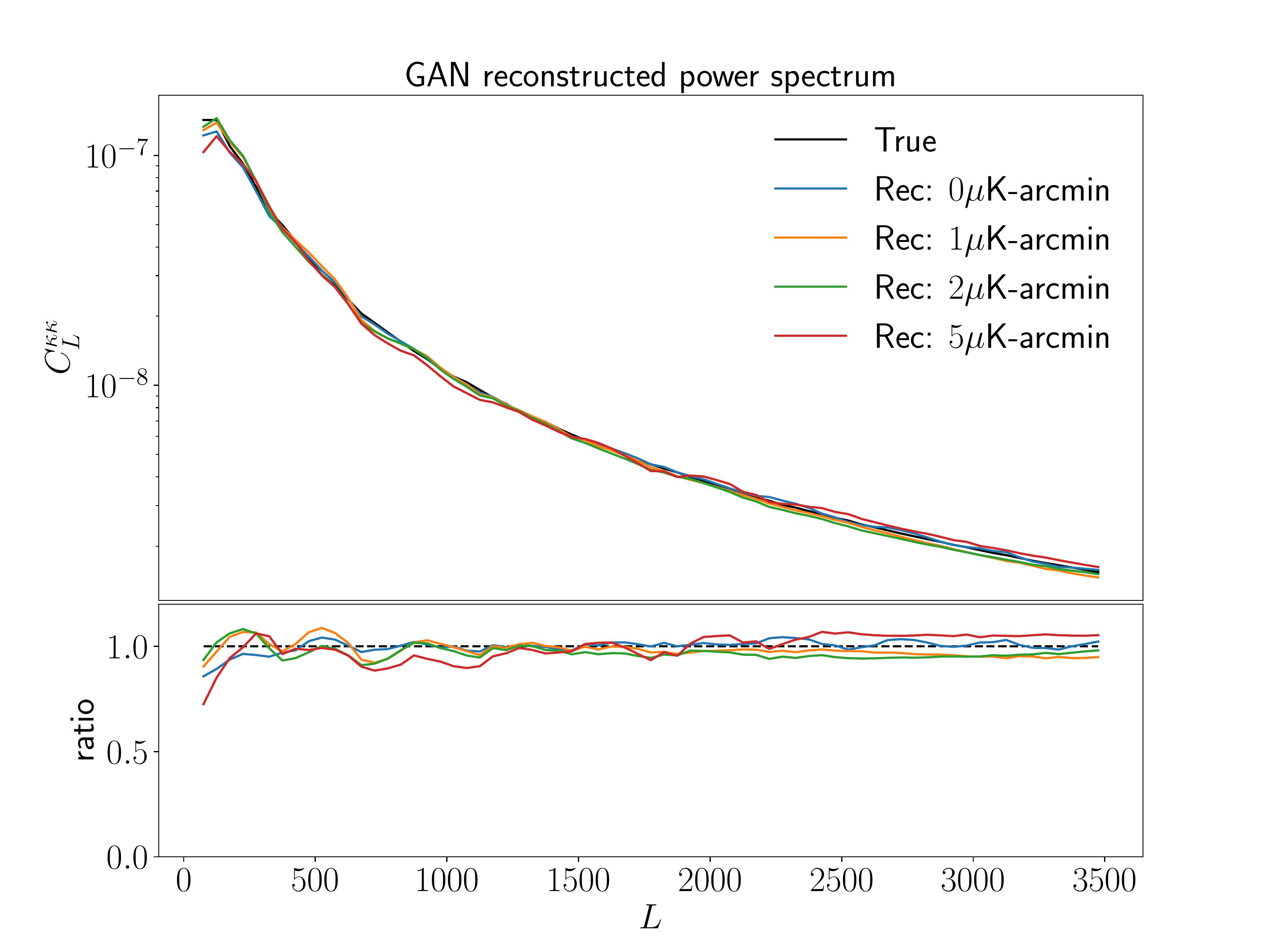}
    \caption{True CMB convergence power spectrum versus reconstructed power spectrum using two deep learning models. The upper half corresponds to the ResUNet, and the lower half is the result of the GAN model. The spectrums are averaged over all realizations in the test set. The ratio of the reconstructed power spectrum to the true power spectrum is also shown here. The length of each $L$ bin is set to be $50$ here and after in this work.}
    \label{fig:mean_ps}
   \end{figure}

As indicated in the previous subsection \ref{sec:level6p2}, ResUNet struggles in reconstructing the small scale structure of the CMB convergence map. This will result in a deficit in the high $L$ part of the convergence power spectrum, as shown in the upper half of Figure~\ref{fig:mean_ps}. The power spectrum plotted is the mean over all test set true/predicted maps. Therefore although our dataset is generated by convergence power spectrum with various scales $\alpha C_{L}^{\kappa\kappa}$ ($0.75<\alpha<1.25$), this figure (Figure~\ref{fig:mean_ps}) still almost perfectly captures the performance of the model on a fixed power spectrum dataset, the same as in \cite{Caldeira:2018ojb}\cite{Guzman:2021nfk}. The ResUNet reconstructed power spectrums, except for the noiseless case, all differ from the true convergence power spectrums substantially. This deficit becomes even more pronounced when $L$ gets larger.

The GAN reconstructed power spectrum can be found in the lower half of Figure~\ref{fig:mean_ps}. The improvement from the GAN model is its successful reconstruction of the convergence power spectrum, even in the presence of detector noise. The credit goes to the design of the discriminator, where the reconstruction of small scale features is emphasized.

\subsection{Training Results: Signal-to-Noise Ratio}
   \begin{figure}
     \centering
        \includegraphics[width=8cm,height=10cm]{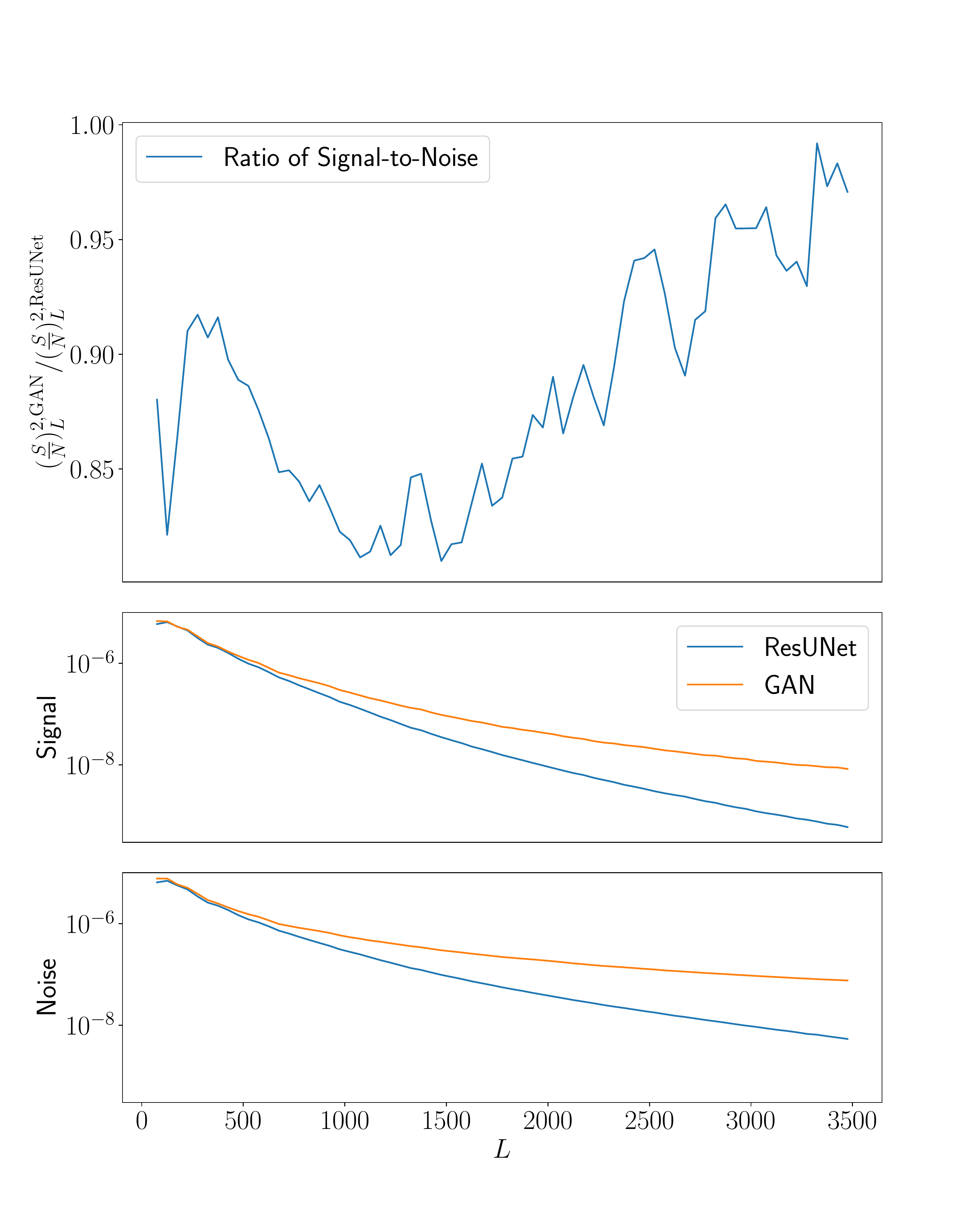}
    \caption{The upper panel is the ratio of the averaged signal-to-noise for the predicted maps on the test set for the two different models. The middle and the bottom panel show the curve for the signal ($C_L^{\kappa\hat\kappa}$) and the noise ($\sqrt{2C_L^{\kappa\kappa}C_L^{\hat\kappa\hat\kappa}/n_L}$) averaged over all realizations, respectively. The detector noise level is taken to be $2\mu$K-armin. Blue and orange correspond the ResUNet's and the GAN's output, respectively.}
    \label{fig:snr}
   \end{figure}
Another valuable metric is the Fourier space signal-to-noise ratio $(S/N)^2_L$. In the flat sky approximation, the signal-to-noise ratio can be expressed as:
\bea 
\bigg(\frac{S}{N}\bigg)^2_L = \frac{n_L(C_L^{\kappa \hat \kappa})^2}{2C_L^{\kappa\kappa}C_L^{\hat\kappa\hat\kappa}}
\eea 
where $C_L^{\kappa \hat \kappa}$ is the cross power spectrum and $n_L$ is the number of modes in each $L$-bin. In Figure~\ref{fig:snr} we show the ratio of the signal-to-noise ratio between the two models, along with the signal and the noise term for the detector noise $2\mu$K-arcmin (all noise levels perform similarly). As shown in the figure, ResUNet has a slightly higher signal-to-noise ratio than the GAN model. This is consistent with the result of subsection \ref{sec:level6p2}: the GAN model is not only putting in signal but also noise. This statement is now proved to be true as in the bottom two panels of Figure~\ref{fig:snr}. We see that when we consider small scales (when $L$ is high), the signal $C_L^{\kappa\hat\kappa}$ is higher for our GAN model, while in the meantime the noise term $\sqrt{2C_L^{\kappa\kappa}C_L^{\hat\kappa\hat\kappa}/n_L}$ is also higher, since the GAN reconstructed power spectrum $C_L^{\hat\kappa\hat\kappa}$ has much more small scale power than the ResUNet one (Figure~\ref{fig:mean_ps}). 

\begin{figure}
     \centering
        \includegraphics[width=8cm,height=6cm]{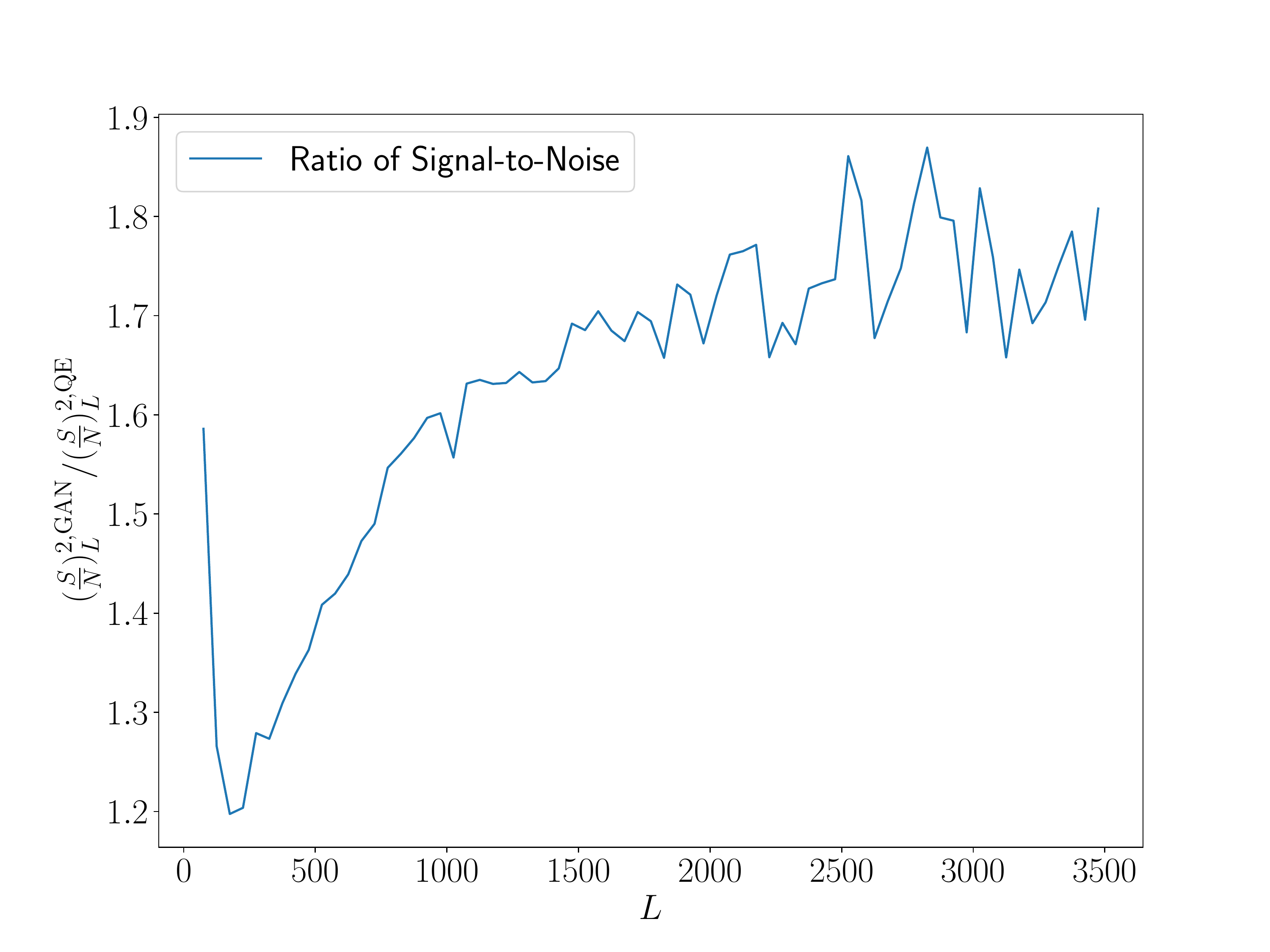}
    \caption{The ratio of the signal-to-noise between the GAN model and the $EB$ quadratic estimator, the detector noise is set to be $2\mu$K-arcmin.}
    \label{fig:snr_qe}
   \end{figure}

Moreover we compare the signal-to-noise ratio between the GAN model and the $EB$ quadratic estimator (QE)\footnote{Notice the performance of the $EB$ QE is almost the minimum variance QE when we only consider polarization maps.} in Figure~\ref{fig:snr_qe} under the $2\mu$K-arcmin detector noise. We can see that although the signal-to-noise ratio of the GAN model is slightly lower than the ResUNet, it is still decently higher than the vanilla QE.

\subsection{Training Results: Robustness Test}
   
\begin{widetext}
   \begin{figure*}
     \centering
        \includegraphics[width=12cm,height=18cm]{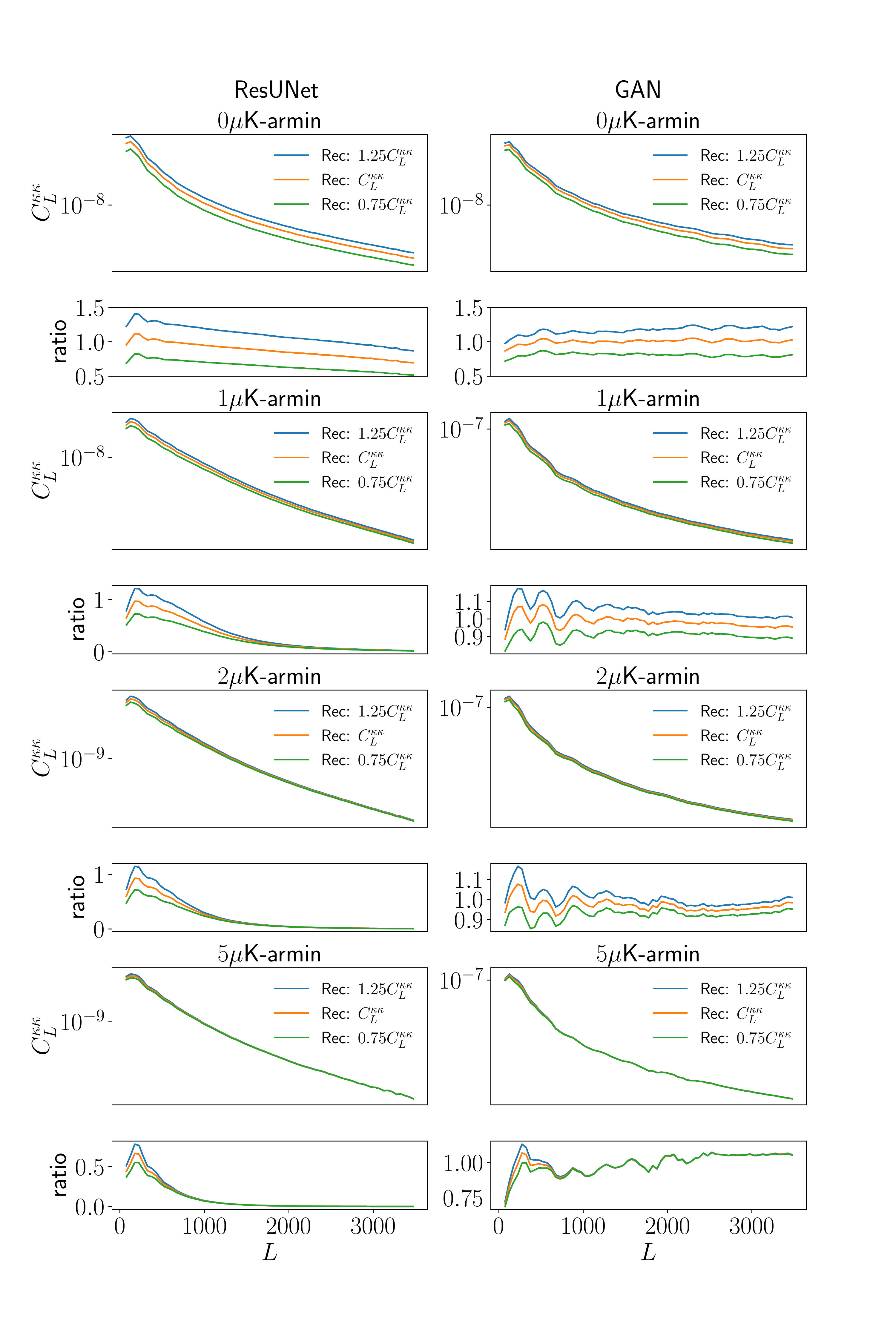}
    \caption{Reconstructed power spectrum averaged over all realizations of the robustness test set's output. Blue, orange and green curves correspond to the reconstructed power spectrums with the corresponding target true convergence maps having power spectrum $1.25C_L^{\kappa\kappa}$, $C_L^{\kappa\kappa}$ and $0.75C_L^{\kappa\kappa}$, respectively. The ratio in this plot is taken to be $C_L^{\hat \kappa\hat \kappa}/C_L^{\kappa\kappa}$, where $C_L^{\hat \kappa\hat \kappa}$ is the predicted power spectrum for both the models.}
    \label{fig:robust}
   \end{figure*}
\end{widetext}

The next thing that we care about most is the generalization capability of the GAN model. Therefore we further test the two models with the following subsequently generated test data: $1,500$ realizations of observed $Q$, $U$ fields along with their associated convergence $\kappa$ maps, created with the convergence power spectrum to be $1.25C_{L}^{\kappa\kappa}$, $C_{L}^{\kappa\kappa}$ and $0.75C_{L}^{\kappa\kappa}$ respectively. We repeat the map generating process for all four different noise levels, and the predicted power spectrum for all these robustness test set using two different models can be found in Figure~\ref{fig:robust}.

The two columns of Figure~\ref{fig:robust} correspond to the ResUNet output and GAN output respectively. We can see that for the noiseless case, both the models are capable of distinguishing the differences in the dataset decently. For the case where detector noise being $1$ or $2\mu$K-arcmin, the performance of both models starts to decrease. However it is still possible for us to train the GAN model iteratively to get an ideal result. Both the models struggle with telling the small scale differences for the highest noise level we consider in this work. 

\section{\label{sec:level7}Conclusion}
In this work we introduce a super-resolution based deep learning architecture designed to reconstruct the CMB lensing potential and compare its performance with the Residual-UNet architecture from previous work. ResUNet targets at minimizing the difference between the generated and the true maps, while the GAN model we apply focuses on reproducing small scale structure. 

We see that these two models each has its own merits: ResUNet is better at reconstructing maps with higher correlations with the true map, while the GAN model is capable of capturing some small scale features and predicting the power spectrum. For example, $2\mu K$-arcmin detector noise, the GAN model correctly predicts the small-scale power spectrum and sacrifices only $\lesssim 15 \%$ signal-to-noise. Notice that the GAN model still has a higher signal-to-noise ratio than the vanilla quadratic estimator's approach. The differences between the two models can also be explained by the ``no free lunch" (NFL) theorem \cite{585893} in the field of machine learning, where it states all optimization algorithms perform equally well when their performance is averaged over all possible objective functions. 

In future work we plan to apply the modified $Pix2PixHD$ model to other reconstruction tasks where the small scale structures are missing (e.g., patchy reionization \cite{Guzman:2021nfk} and CMB polarization rotation \cite{Guzman:2021ygf}). Further we plan to include galactic and extra-galactic foreground effects in the input polarization maps, and learn to recover the CMB signal as well as the lensing signal simultaneously. More realistically we plan to consider the full spherical sky instead of a small flat patch, this can be achieved by transforming the spherical map into a two-dimensional array with the NESTED scheme of HEALPix \cite{Gorski:2004by}\cite{Wang:2022ybb}. In terms of the lensing reconstruction task itself, the main challenge right now for these deep learning models is to correctly recover the convergence power spectrum when trained on a variety of spectra, none of which may equal the true spectrum. We are exploring the possibility that a more ``flexible'' and ``physical'' architecture -- the variational autoencoder (VAE) \cite{Kingma2014} -- will potentially outperform the GAN model in this respect. The VAE based super-resolution architecture is still under active development \cite{2020SPIE11400E..0UH}\cite{https://doi.org/10.48550/arxiv.2203.09445}.

\clearpage
\section*{Acknowledgement}
We thank Yueying Ni and Eric Baxter for useful discussions. This work is supported by U.S. Dept. of Energy contract DE-SC0019248 and NSF AST-1909193.


\begin{thebibliography}{35}%
\makeatletter
\providecommand \@ifxundefined [1]{%
 \@ifx{#1\undefined}
}%
\providecommand \@ifnum [1]{%
 \ifnum #1\expandafter \@firstoftwo
 \else \expandafter \@secondoftwo
 \fi
}%
\providecommand \@ifx [1]{%
 \ifx #1\expandafter \@firstoftwo
 \else \expandafter \@secondoftwo
 \fi
}%
\providecommand \natexlab [1]{#1}%
\providecommand \enquote  [1]{``#1''}%
\providecommand \bibnamefont  [1]{#1}%
\providecommand \bibfnamefont [1]{#1}%
\providecommand \citenamefont [1]{#1}%
\providecommand \href@noop [0]{\@secondoftwo}%
\providecommand \href [0]{\begingroup \@sanitize@url \@href}%
\providecommand \@href[1]{\@@startlink{#1}\@@href}%
\providecommand \@@href[1]{\endgroup#1\@@endlink}%
\providecommand \@sanitize@url [0]{\catcode `\\12\catcode `\$12\catcode
  `\&12\catcode `\#12\catcode `\^12\catcode `\_12\catcode `\%12\relax}%
\providecommand \@@startlink[1]{}%
\providecommand \@@endlink[0]{}%
\providecommand \url  [0]{\begingroup\@sanitize@url \@url }%
\providecommand \@url [1]{\endgroup\@href {#1}{\urlprefix }}%
\providecommand \urlprefix  [0]{URL }%
\providecommand \Eprint [0]{\href }%
\providecommand \doibase [0]{http://dx.doi.org/}%
\providecommand \selectlanguage [0]{\@gobble}%
\providecommand \bibinfo  [0]{\@secondoftwo}%
\providecommand \bibfield  [0]{\@secondoftwo}%
\providecommand \translation [1]{[#1]}%
\providecommand \BibitemOpen [0]{}%
\providecommand \bibitemStop [0]{}%
\providecommand \bibitemNoStop [0]{.\EOS\space}%
\providecommand \EOS [0]{\spacefactor3000\relax}%
\providecommand \BibitemShut  [1]{\csname bibitem#1\endcsname}%
\let\auto@bib@innerbib\@empty
\bibitem [{\citenamefont {Hu}(2008)}]{Hu:2008hd}%
  \BibitemOpen
  \bibfield  {author} {\bibinfo {author} {\bibfnamefont {W.}~\bibnamefont
  {Hu}},\ }\href@noop {} {\  (\bibinfo {year} {2008})},\ \Eprint
  {http://arxiv.org/abs/0802.3688} {arXiv:0802.3688 [astro-ph]} \BibitemShut
  {NoStop}%
\bibitem [{\citenamefont {Lewis}\ and\ \citenamefont
  {Challinor}(2006)}]{Lewis:2006fu}%
  \BibitemOpen
  \bibfield  {author} {\bibinfo {author} {\bibfnamefont {A.}~\bibnamefont
  {Lewis}}\ and\ \bibinfo {author} {\bibfnamefont {A.}~\bibnamefont
  {Challinor}},\ }\href {\doibase 10.1016/j.physrep.2006.03.002} {\bibfield
  {journal} {\bibinfo  {journal} {Phys. Rept.}\ }\textbf {\bibinfo {volume}
  {429}},\ \bibinfo {pages} {1} (\bibinfo {year} {2006})},\ \Eprint
  {http://arxiv.org/abs/astro-ph/0601594} {arXiv:astro-ph/0601594} \BibitemShut
  {NoStop}%
\bibitem [{\citenamefont {Aghanim}\ \emph {et~al.}(2020)\citenamefont {Aghanim}
  \emph {et~al.}}]{Planck:2018lbu}%
  \BibitemOpen
  \bibfield  {author} {\bibinfo {author} {\bibfnamefont {N.}~\bibnamefont
  {Aghanim}} \emph {et~al.} (\bibinfo {collaboration} {Planck}),\ }\href
  {\doibase 10.1051/0004-6361/201833886} {\bibfield  {journal} {\bibinfo
  {journal} {Astron. Astrophys.}\ }\textbf {\bibinfo {volume} {641}},\ \bibinfo
  {pages} {A8} (\bibinfo {year} {2020})},\ \Eprint
  {http://arxiv.org/abs/1807.06210} {arXiv:1807.06210 [astro-ph.CO]}
  \BibitemShut {NoStop}%
\bibitem [{\citenamefont {Hu}(2001)}]{Hu:2001tn}%
  \BibitemOpen
  \bibfield  {author} {\bibinfo {author} {\bibfnamefont {W.}~\bibnamefont
  {Hu}},\ }\href {\doibase 10.1086/323253} {\bibfield  {journal} {\bibinfo
  {journal} {Astrophys. J. Lett.}\ }\textbf {\bibinfo {volume} {557}},\
  \bibinfo {pages} {L79} (\bibinfo {year} {2001})},\ \Eprint
  {http://arxiv.org/abs/astro-ph/0105424} {arXiv:astro-ph/0105424} \BibitemShut
  {NoStop}%
\bibitem [{\citenamefont {Henderson}\ \emph {et~al.}(2016)\citenamefont
  {Henderson} \emph {et~al.}}]{Henderson:2015nzj}%
  \BibitemOpen
  \bibfield  {author} {\bibinfo {author} {\bibfnamefont {S.~W.}\ \bibnamefont
  {Henderson}} \emph {et~al.},\ }\href {\doibase 10.1007/s10909-016-1575-z}
  {\bibfield  {journal} {\bibinfo  {journal} {J. Low Temp. Phys.}\ }\textbf
  {\bibinfo {volume} {184}},\ \bibinfo {pages} {772} (\bibinfo {year}
  {2016})},\ \Eprint {http://arxiv.org/abs/1510.02809} {arXiv:1510.02809
  [astro-ph.IM]} \BibitemShut {NoStop}%
\bibitem [{\citenamefont {Benson}\ \emph {et~al.}(2014)\citenamefont {Benson}
  \emph {et~al.}}]{SPT-3G:2014dbx}%
  \BibitemOpen
  \bibfield  {author} {\bibinfo {author} {\bibfnamefont {B.~A.}\ \bibnamefont
  {Benson}} \emph {et~al.} (\bibinfo {collaboration} {SPT-3G}),\ }\href
  {\doibase 10.1117/12.2057305} {\bibfield  {journal} {\bibinfo  {journal}
  {Proc. SPIE Int. Soc. Opt. Eng.}\ }\textbf {\bibinfo {volume} {9153}},\
  \bibinfo {pages} {91531P} (\bibinfo {year} {2014})},\ \Eprint
  {http://arxiv.org/abs/1407.2973} {arXiv:1407.2973 [astro-ph.IM]} \BibitemShut
  {NoStop}%
\bibitem [{\citenamefont {Ade}\ \emph {et~al.}(2019)\citenamefont {Ade} \emph
  {et~al.}}]{SimonsObservatory:2018koc}%
  \BibitemOpen
  \bibfield  {author} {\bibinfo {author} {\bibfnamefont {P.}~\bibnamefont
  {Ade}} \emph {et~al.} (\bibinfo {collaboration} {Simons Observatory}),\
  }\href {\doibase 10.1088/1475-7516/2019/02/056} {\bibfield  {journal}
  {\bibinfo  {journal} {JCAP}\ }\textbf {\bibinfo {volume} {02}},\ \bibinfo
  {pages} {056} (\bibinfo {year} {2019})},\ \Eprint
  {http://arxiv.org/abs/1808.07445} {arXiv:1808.07445 [astro-ph.CO]}
  \BibitemShut {NoStop}%
\bibitem [{\citenamefont {Hui}\ \emph {et~al.}(2018)\citenamefont {Hui} \emph
  {et~al.}}]{Hui:2018cvg}%
  \BibitemOpen
  \bibfield  {author} {\bibinfo {author} {\bibfnamefont {H.}~\bibnamefont
  {Hui}} \emph {et~al.},\ }\href {\doibase 10.1117/12.2311725} {\bibfield
  {journal} {\bibinfo  {journal} {Proc. SPIE Int. Soc. Opt. Eng.}\ }\textbf
  {\bibinfo {volume} {10708}},\ \bibinfo {pages} {1070807} (\bibinfo {year}
  {2018})},\ \Eprint {http://arxiv.org/abs/1808.00568} {arXiv:1808.00568
  [astro-ph.IM]} \BibitemShut {NoStop}%
\bibitem [{\citenamefont {Aravena}\ \emph {et~al.}(2019)\citenamefont {Aravena}
  \emph {et~al.}}]{Aravena:2019tye}%
  \BibitemOpen
  \bibfield  {author} {\bibinfo {author} {\bibfnamefont {M.}~\bibnamefont
  {Aravena}} \emph {et~al.},\ }\href@noop {} {\  (\bibinfo {year} {2019})},\
  \Eprint {http://arxiv.org/abs/1909.02587} {arXiv:1909.02587 [astro-ph.IM]}
  \BibitemShut {NoStop}%
\bibitem [{\citenamefont {Abazajian}\ \emph {et~al.}(2016)\citenamefont
  {Abazajian} \emph {et~al.}}]{CMB-S4:2016ple}%
  \BibitemOpen
  \bibfield  {author} {\bibinfo {author} {\bibfnamefont {K.~N.}\ \bibnamefont
  {Abazajian}} \emph {et~al.} (\bibinfo {collaboration} {CMB-S4}),\ }\href@noop
  {} {\  (\bibinfo {year} {2016})},\ \Eprint {http://arxiv.org/abs/1610.02743}
  {arXiv:1610.02743 [astro-ph.CO]} \BibitemShut {NoStop}%
\bibitem [{\citenamefont {Hanany}\ \emph {et~al.}(2019)\citenamefont {Hanany}
  \emph {et~al.}}]{NASAPICO:2019thw}%
  \BibitemOpen
  \bibfield  {author} {\bibinfo {author} {\bibfnamefont {S.}~\bibnamefont
  {Hanany}} \emph {et~al.} (\bibinfo {collaboration} {NASA PICO}),\ }\href@noop
  {} {\  (\bibinfo {year} {2019})},\ \Eprint {http://arxiv.org/abs/1902.10541}
  {arXiv:1902.10541 [astro-ph.IM]} \BibitemShut {NoStop}%
\bibitem [{\citenamefont {Maniyar}\ \emph {et~al.}(2021)\citenamefont
  {Maniyar}, \citenamefont {Ali-Ha\"\i{}moud}, \citenamefont {Carron},
  \citenamefont {Lewis},\ and\ \citenamefont
  {Madhavacheril}}]{Maniyar:2021msb}%
  \BibitemOpen
  \bibfield  {author} {\bibinfo {author} {\bibfnamefont {A.~S.}\ \bibnamefont
  {Maniyar}}, \bibinfo {author} {\bibfnamefont {Y.}~\bibnamefont
  {Ali-Ha\"\i{}moud}}, \bibinfo {author} {\bibfnamefont {J.}~\bibnamefont
  {Carron}}, \bibinfo {author} {\bibfnamefont {A.}~\bibnamefont {Lewis}}, \
  and\ \bibinfo {author} {\bibfnamefont {M.~S.}\ \bibnamefont
  {Madhavacheril}},\ }\href {\doibase 10.1103/PhysRevD.103.083524} {\bibfield
  {journal} {\bibinfo  {journal} {Phys. Rev. D}\ }\textbf {\bibinfo {volume}
  {103}},\ \bibinfo {pages} {083524} (\bibinfo {year} {2021})},\ \Eprint
  {http://arxiv.org/abs/2101.12193} {arXiv:2101.12193 [astro-ph.CO]}
  \BibitemShut {NoStop}%
\bibitem [{\citenamefont {Hirata}\ and\ \citenamefont
  {Seljak}(2003)}]{Hirata:2003ka}%
  \BibitemOpen
  \bibfield  {author} {\bibinfo {author} {\bibfnamefont {C.~M.}\ \bibnamefont
  {Hirata}}\ and\ \bibinfo {author} {\bibfnamefont {U.}~\bibnamefont
  {Seljak}},\ }\href {\doibase 10.1103/PhysRevD.68.083002} {\bibfield
  {journal} {\bibinfo  {journal} {Phys. Rev. D}\ }\textbf {\bibinfo {volume}
  {68}},\ \bibinfo {pages} {083002} (\bibinfo {year} {2003})},\ \Eprint
  {http://arxiv.org/abs/astro-ph/0306354} {arXiv:astro-ph/0306354} \BibitemShut
  {NoStop}%
\bibitem [{\citenamefont {{Smith}}\ \emph {et~al.}(2012)\citenamefont
  {{Smith}}, \citenamefont {{Hanson}}, \citenamefont {{LoVerde}}, \citenamefont
  {{Hirata}},\ and\ \citenamefont {{Zahn}}}]{2012JCAP...06..014S}%
  \BibitemOpen
  \bibfield  {author} {\bibinfo {author} {\bibfnamefont {K.~M.}\ \bibnamefont
  {{Smith}}}, \bibinfo {author} {\bibfnamefont {D.}~\bibnamefont {{Hanson}}},
  \bibinfo {author} {\bibfnamefont {M.}~\bibnamefont {{LoVerde}}}, \bibinfo
  {author} {\bibfnamefont {C.~M.}\ \bibnamefont {{Hirata}}}, \ and\ \bibinfo
  {author} {\bibfnamefont {O.}~\bibnamefont {{Zahn}}},\ }\href {\doibase
  10.1088/1475-7516/2012/06/014} {\bibfield  {journal} {\bibinfo  {journal} {J.
  Cosmol. Astropart. Phys.}\ }\textbf {\bibinfo {volume} {2012}},\ \bibinfo
  {eid} {014} (\bibinfo {year} {2012})},\ \Eprint
  {http://arxiv.org/abs/1010.0048} {arXiv:1010.0048 [astro-ph.CO]} \BibitemShut
  {NoStop}%
\bibitem [{\citenamefont {Hadzhiyska}\ \emph {et~al.}(2019)\citenamefont
  {Hadzhiyska}, \citenamefont {Sherwin}, \citenamefont {Madhavacheril},\ and\
  \citenamefont {Ferraro}}]{Hadzhiyska:2019cle}%
  \BibitemOpen
  \bibfield  {author} {\bibinfo {author} {\bibfnamefont {B.}~\bibnamefont
  {Hadzhiyska}}, \bibinfo {author} {\bibfnamefont {B.~D.}\ \bibnamefont
  {Sherwin}}, \bibinfo {author} {\bibfnamefont {M.}~\bibnamefont
  {Madhavacheril}}, \ and\ \bibinfo {author} {\bibfnamefont {S.}~\bibnamefont
  {Ferraro}},\ }\href {\doibase 10.1103/PhysRevD.100.023547} {\bibfield
  {journal} {\bibinfo  {journal} {Phys. Rev. D}\ }\textbf {\bibinfo {volume}
  {100}},\ \bibinfo {pages} {023547} (\bibinfo {year} {2019})},\ \Eprint
  {http://arxiv.org/abs/1905.04217} {arXiv:1905.04217 [astro-ph.CO]}
  \BibitemShut {NoStop}%
\bibitem [{\citenamefont {Carron}\ and\ \citenamefont
  {Lewis}(2017)}]{Carron:2017mqf}%
  \BibitemOpen
  \bibfield  {author} {\bibinfo {author} {\bibfnamefont {J.}~\bibnamefont
  {Carron}}\ and\ \bibinfo {author} {\bibfnamefont {A.}~\bibnamefont {Lewis}},\
  }\href {\doibase 10.1103/PhysRevD.96.063510} {\bibfield  {journal} {\bibinfo
  {journal} {Phys. Rev. D}\ }\textbf {\bibinfo {volume} {96}},\ \bibinfo
  {pages} {063510} (\bibinfo {year} {2017})},\ \Eprint
  {http://arxiv.org/abs/1704.08230} {arXiv:1704.08230 [astro-ph.CO]}
  \BibitemShut {NoStop}%
\bibitem [{\citenamefont {Millea}\ \emph {et~al.}(2019)\citenamefont {Millea},
  \citenamefont {Anderes},\ and\ \citenamefont {Wandelt}}]{Millea:2017fyd}%
  \BibitemOpen
  \bibfield  {author} {\bibinfo {author} {\bibfnamefont {M.}~\bibnamefont
  {Millea}}, \bibinfo {author} {\bibfnamefont {E.}~\bibnamefont {Anderes}}, \
  and\ \bibinfo {author} {\bibfnamefont {B.~D.}\ \bibnamefont {Wandelt}},\
  }\href {\doibase 10.1103/PhysRevD.100.023509} {\bibfield  {journal} {\bibinfo
   {journal} {Phys. Rev. D}\ }\textbf {\bibinfo {volume} {100}},\ \bibinfo
  {pages} {023509} (\bibinfo {year} {2019})},\ \Eprint
  {http://arxiv.org/abs/1708.06753} {arXiv:1708.06753 [astro-ph.CO]}
  \BibitemShut {NoStop}%
\bibitem [{\citenamefont {Caldeira}\ \emph {et~al.}(2019)\citenamefont
  {Caldeira}, \citenamefont {Wu}, \citenamefont {Nord}, \citenamefont
  {Avestruz}, \citenamefont {Trivedi},\ and\ \citenamefont
  {Story}}]{Caldeira:2018ojb}%
  \BibitemOpen
  \bibfield  {author} {\bibinfo {author} {\bibfnamefont {J.~a.}\ \bibnamefont
  {Caldeira}}, \bibinfo {author} {\bibfnamefont {W.~L.~K.}\ \bibnamefont {Wu}},
  \bibinfo {author} {\bibfnamefont {B.}~\bibnamefont {Nord}}, \bibinfo {author}
  {\bibfnamefont {C.}~\bibnamefont {Avestruz}}, \bibinfo {author}
  {\bibfnamefont {S.}~\bibnamefont {Trivedi}}, \ and\ \bibinfo {author}
  {\bibfnamefont {K.~T.}\ \bibnamefont {Story}},\ }\href {\doibase
  10.1016/j.ascom.2019.100307} {\bibfield  {journal} {\bibinfo  {journal}
  {Astron. Comput.}\ }\textbf {\bibinfo {volume} {28}},\ \bibinfo {pages}
  {100307} (\bibinfo {year} {2019})},\ \Eprint
  {http://arxiv.org/abs/1810.01483} {arXiv:1810.01483 [astro-ph.CO]}
  \BibitemShut {NoStop}%
\bibitem [{\citenamefont {Guzman}\ and\ \citenamefont
  {Meyers}(2021)}]{Guzman:2021nfk}%
  \BibitemOpen
  \bibfield  {author} {\bibinfo {author} {\bibfnamefont {E.}~\bibnamefont
  {Guzman}}\ and\ \bibinfo {author} {\bibfnamefont {J.}~\bibnamefont
  {Meyers}},\ }\href {\doibase 10.1103/PhysRevD.104.043529} {\bibfield
  {journal} {\bibinfo  {journal} {Phys. Rev. D}\ }\textbf {\bibinfo {volume}
  {104}},\ \bibinfo {pages} {043529} (\bibinfo {year} {2021})},\ \Eprint
  {http://arxiv.org/abs/2101.01214} {arXiv:2101.01214 [astro-ph.CO]}
  \BibitemShut {NoStop}%
\bibitem [{\citenamefont {{Kayalibay}}\ \emph {et~al.}(2017)\citenamefont
  {{Kayalibay}}, \citenamefont {{Jensen}},\ and\ \citenamefont {{van der
  Smagt}}}]{2017arXiv170103056K}%
  \BibitemOpen
  \bibfield  {author} {\bibinfo {author} {\bibfnamefont {B.}~\bibnamefont
  {{Kayalibay}}}, \bibinfo {author} {\bibfnamefont {G.}~\bibnamefont
  {{Jensen}}}, \ and\ \bibinfo {author} {\bibfnamefont {P.}~\bibnamefont {{van
  der Smagt}}},\ }\href@noop {} {\bibfield  {journal} {\bibinfo  {journal}
  {arXiv e-prints}\ ,\ \bibinfo {eid} {arXiv:1701.03056}} (\bibinfo {year}
  {2017})},\ \Eprint {http://arxiv.org/abs/1701.03056} {arXiv:1701.03056
  [cs.CV]} \BibitemShut {NoStop}%
\bibitem [{\citenamefont {{Zhang}}\ \emph {et~al.}(2018)\citenamefont
  {{Zhang}}, \citenamefont {{Liu}},\ and\ \citenamefont
  {{Wang}}}]{2018IGRSL..15..749Z}%
  \BibitemOpen
  \bibfield  {author} {\bibinfo {author} {\bibfnamefont {Z.}~\bibnamefont
  {{Zhang}}}, \bibinfo {author} {\bibfnamefont {Q.}~\bibnamefont {{Liu}}}, \
  and\ \bibinfo {author} {\bibfnamefont {Y.}~\bibnamefont {{Wang}}},\ }\href
  {\doibase 10.1109/LGRS.2018.2802944} {\bibfield  {journal} {\bibinfo
  {journal} {IEEE Geoscience and Remote Sensing Letters}\ }\textbf {\bibinfo
  {volume} {15}},\ \bibinfo {pages} {749} (\bibinfo {year} {2018})},\ \Eprint
  {http://arxiv.org/abs/1711.10684} {arXiv:1711.10684 [cs.CV]} \BibitemShut
  {NoStop}%
\bibitem [{\citenamefont {{Wang}}\ \emph {et~al.}(2017)\citenamefont {{Wang}},
  \citenamefont {{Liu}}, \citenamefont {{Zhu}}, \citenamefont {{Tao}},
  \citenamefont {{Kautz}},\ and\ \citenamefont
  {{Catanzaro}}}]{2017arXiv171111585W}%
  \BibitemOpen
  \bibfield  {author} {\bibinfo {author} {\bibfnamefont {T.-C.}\ \bibnamefont
  {{Wang}}}, \bibinfo {author} {\bibfnamefont {M.-Y.}\ \bibnamefont {{Liu}}},
  \bibinfo {author} {\bibfnamefont {J.-Y.}\ \bibnamefont {{Zhu}}}, \bibinfo
  {author} {\bibfnamefont {A.}~\bibnamefont {{Tao}}}, \bibinfo {author}
  {\bibfnamefont {J.}~\bibnamefont {{Kautz}}}, \ and\ \bibinfo {author}
  {\bibfnamefont {B.}~\bibnamefont {{Catanzaro}}},\ }\href@noop {} {\bibfield
  {journal} {\bibinfo  {journal} {arXiv e-prints}\ ,\ \bibinfo {eid}
  {arXiv:1711.11585}} (\bibinfo {year} {2017})},\ \Eprint
  {http://arxiv.org/abs/1711.11585} {arXiv:1711.11585 [cs.CV]} \BibitemShut
  {NoStop}%
\bibitem [{\citenamefont {{Mirza}}\ and\ \citenamefont
  {{Osindero}}(2014)}]{2014arXiv1411.1784M}%
  \BibitemOpen
  \bibfield  {author} {\bibinfo {author} {\bibfnamefont {M.}~\bibnamefont
  {{Mirza}}}\ and\ \bibinfo {author} {\bibfnamefont {S.}~\bibnamefont
  {{Osindero}}},\ }\href@noop {} {\bibfield  {journal} {\bibinfo  {journal}
  {arXiv e-prints}\ ,\ \bibinfo {eid} {arXiv:1411.1784}} (\bibinfo {year}
  {2014})},\ \Eprint {http://arxiv.org/abs/1411.1784} {arXiv:1411.1784 [cs.LG]}
  \BibitemShut {NoStop}%
\bibitem [{\citenamefont {Knox}(1995)}]{PhysRevD.52.4307}%
  \BibitemOpen
  \bibfield  {author} {\bibinfo {author} {\bibfnamefont {L.}~\bibnamefont
  {Knox}},\ }\href {\doibase 10.1103/PhysRevD.52.4307} {\bibfield  {journal}
  {\bibinfo  {journal} {Phys. Rev. D}\ }\textbf {\bibinfo {volume} {52}},\
  \bibinfo {pages} {4307} (\bibinfo {year} {1995})}\BibitemShut {NoStop}%
\bibitem [{\citenamefont {Zaldarriaga}(2001)}]{Zaldarriaga:2001st}%
  \BibitemOpen
  \bibfield  {author} {\bibinfo {author} {\bibfnamefont {M.}~\bibnamefont
  {Zaldarriaga}},\ }\href {\doibase 10.1103/PhysRevD.64.103001} {\bibfield
  {journal} {\bibinfo  {journal} {Phys. Rev. D}\ }\textbf {\bibinfo {volume}
  {64}},\ \bibinfo {pages} {103001} (\bibinfo {year} {2001})},\ \Eprint
  {http://arxiv.org/abs/astro-ph/0106174} {arXiv:astro-ph/0106174} \BibitemShut
  {NoStop}%
\bibitem [{\citenamefont {{Shin}}\ \emph {et~al.}(2020)\citenamefont {{Shin}},
  \citenamefont {{Moon}}, \citenamefont {{Park}}, \citenamefont {{Jeong}},
  \citenamefont {{Lee}},\ and\ \citenamefont {{Bae}}}]{2020ApJ...895L..16S}%
  \BibitemOpen
  \bibfield  {author} {\bibinfo {author} {\bibfnamefont {G.}~\bibnamefont
  {{Shin}}}, \bibinfo {author} {\bibfnamefont {Y.-J.}\ \bibnamefont {{Moon}}},
  \bibinfo {author} {\bibfnamefont {E.}~\bibnamefont {{Park}}}, \bibinfo
  {author} {\bibfnamefont {H.}~\bibnamefont {{Jeong}}}, \bibinfo {author}
  {\bibfnamefont {H.}~\bibnamefont {{Lee}}}, \ and\ \bibinfo {author}
  {\bibfnamefont {S.-H.}\ \bibnamefont {{Bae}}},\ }\href {\doibase
  10.3847/2041-8213/ab9085} {\bibfield  {journal} {\bibinfo  {journal}
  {Astrophys. J. Lett.}\ }\textbf {\bibinfo {volume} {895}},\ \bibinfo {eid}
  {L16} (\bibinfo {year} {2020})}\BibitemShut {NoStop}%
\bibitem [{\citenamefont {Hu}(2000)}]{Hu:1999vq}%
  \BibitemOpen
  \bibfield  {author} {\bibinfo {author} {\bibfnamefont {W.}~\bibnamefont
  {Hu}},\ }\href {\doibase 10.1086/308279} {\bibfield  {journal} {\bibinfo
  {journal} {Astrophys. J.}\ }\textbf {\bibinfo {volume} {529}},\ \bibinfo
  {pages} {12} (\bibinfo {year} {2000})},\ \Eprint
  {http://arxiv.org/abs/astro-ph/9907103} {arXiv:astro-ph/9907103} \BibitemShut
  {NoStop}%
\bibitem [{\citenamefont {{Kingma}}\ and\ \citenamefont
  {{Ba}}(2014)}]{2014arXiv1412.6980K}%
  \BibitemOpen
  \bibfield  {author} {\bibinfo {author} {\bibfnamefont {D.~P.}\ \bibnamefont
  {{Kingma}}}\ and\ \bibinfo {author} {\bibfnamefont {J.}~\bibnamefont
  {{Ba}}},\ }\href@noop {} {\bibfield  {journal} {\bibinfo  {journal} {arXiv
  e-prints}\ ,\ \bibinfo {eid} {arXiv:1412.6980}} (\bibinfo {year} {2014})},\
  \Eprint {http://arxiv.org/abs/1412.6980} {arXiv:1412.6980 [cs.LG]}
  \BibitemShut {NoStop}%
\bibitem [{\citenamefont {Wolpert}\ and\ \citenamefont
  {Macready}(1997)}]{585893}%
  \BibitemOpen
  \bibfield  {author} {\bibinfo {author} {\bibfnamefont {D.}~\bibnamefont
  {Wolpert}}\ and\ \bibinfo {author} {\bibfnamefont {W.}~\bibnamefont
  {Macready}},\ }\href {\doibase 10.1109/4235.585893} {\bibfield  {journal}
  {\bibinfo  {journal} {IEEE Transactions on Evolutionary Computation}\
  }\textbf {\bibinfo {volume} {1}},\ \bibinfo {pages} {67} (\bibinfo {year}
  {1997})}\BibitemShut {NoStop}%
\bibitem [{\citenamefont {Guzman}\ and\ \citenamefont
  {Meyers}(2022)}]{Guzman:2021ygf}%
  \BibitemOpen
  \bibfield  {author} {\bibinfo {author} {\bibfnamefont {E.}~\bibnamefont
  {Guzman}}\ and\ \bibinfo {author} {\bibfnamefont {J.}~\bibnamefont
  {Meyers}},\ }\href {\doibase 10.1088/1475-7516/2022/01/030} {\bibfield
  {journal} {\bibinfo  {journal} {JCAP}\ }\textbf {\bibinfo {volume} {01}},\
  \bibinfo {pages} {030} (\bibinfo {year} {2022})},\ \Eprint
  {http://arxiv.org/abs/2109.09715} {arXiv:2109.09715 [astro-ph.CO]}
  \BibitemShut {NoStop}%
\bibitem [{\citenamefont {G\'orski}\ \emph {et~al.}(2005)\citenamefont
  {G\'orski}, \citenamefont {Hivon}, \citenamefont {Banday}, \citenamefont
  {Wandelt}, \citenamefont {Hansen}, \citenamefont {Reinecke},\ and\
  \citenamefont {Bartelman}}]{Gorski:2004by}%
  \BibitemOpen
  \bibfield  {author} {\bibinfo {author} {\bibfnamefont {K.~M.}\ \bibnamefont
  {G\'orski}}, \bibinfo {author} {\bibfnamefont {E.}~\bibnamefont {Hivon}},
  \bibinfo {author} {\bibfnamefont {A.~J.}\ \bibnamefont {Banday}}, \bibinfo
  {author} {\bibfnamefont {B.~D.}\ \bibnamefont {Wandelt}}, \bibinfo {author}
  {\bibfnamefont {F.~K.}\ \bibnamefont {Hansen}}, \bibinfo {author}
  {\bibfnamefont {M.}~\bibnamefont {Reinecke}}, \ and\ \bibinfo {author}
  {\bibfnamefont {M.}~\bibnamefont {Bartelman}},\ }\href {\doibase
  10.1086/427976} {\bibfield  {journal} {\bibinfo  {journal} {Astrophys. J.}\
  }\textbf {\bibinfo {volume} {622}},\ \bibinfo {pages} {759} (\bibinfo {year}
  {2005})},\ \Eprint {http://arxiv.org/abs/astro-ph/0409513}
  {arXiv:astro-ph/0409513} \BibitemShut {NoStop}%
\bibitem [{\citenamefont {Wang}\ \emph {et~al.}(2022)\citenamefont {Wang},
  \citenamefont {Shi}, \citenamefont {Yan}, \citenamefont {Xia}, \citenamefont
  {Zhao}, \citenamefont {Li},\ and\ \citenamefont {Li}}]{Wang:2022ybb}%
  \BibitemOpen
  \bibfield  {author} {\bibinfo {author} {\bibfnamefont {G.-J.}\ \bibnamefont
  {Wang}}, \bibinfo {author} {\bibfnamefont {H.-L.}\ \bibnamefont {Shi}},
  \bibinfo {author} {\bibfnamefont {Y.-P.}\ \bibnamefont {Yan}}, \bibinfo
  {author} {\bibfnamefont {J.-Q.}\ \bibnamefont {Xia}}, \bibinfo {author}
  {\bibfnamefont {Y.-Y.}\ \bibnamefont {Zhao}}, \bibinfo {author}
  {\bibfnamefont {S.-Y.}\ \bibnamefont {Li}}, \ and\ \bibinfo {author}
  {\bibfnamefont {J.-F.}\ \bibnamefont {Li}},\ }\href {\doibase
  10.3847/1538-4365/ac5f4a} {\bibfield  {journal} {\bibinfo  {journal}
  {Astrophys. J. Supp.}\ }\textbf {\bibinfo {volume} {260}},\ \bibinfo {pages}
  {13} (\bibinfo {year} {2022})},\ \Eprint {http://arxiv.org/abs/2204.01820}
  {arXiv:2204.01820 [astro-ph.CO]} \BibitemShut {NoStop}%
\bibitem [{\citenamefont {Kingma}\ and\ \citenamefont
  {Welling}(2014)}]{Kingma2014}%
  \BibitemOpen
  \bibfield  {author} {\bibinfo {author} {\bibfnamefont {D.~P.}\ \bibnamefont
  {Kingma}}\ and\ \bibinfo {author} {\bibfnamefont {M.}~\bibnamefont
  {Welling}},\ }in\ \href@noop {} {\emph {\bibinfo {booktitle} {2nd
  International Conference on Learning Representations, {ICLR} 2014, Banff, AB,
  Canada, April 14-16, 2014, Conference Track Proceedings}}}\ (\bibinfo {year}
  {2014})\ \Eprint {http://arxiv.org/abs/http://arxiv.org/abs/1312.6114v10}
  {http://arxiv.org/abs/1312.6114v10} \BibitemShut {NoStop}%
\bibitem [{\citenamefont {{Heydari}}\ and\ \citenamefont
  {{Mehmood}}(2020)}]{2020SPIE11400E..0UH}%
  \BibitemOpen
  \bibfield  {author} {\bibinfo {author} {\bibfnamefont {A.~A.}\ \bibnamefont
  {{Heydari}}}\ and\ \bibinfo {author} {\bibfnamefont {A.}~\bibnamefont
  {{Mehmood}}},\ }in\ \href {\doibase 10.1117/12.2559808} {\emph {\bibinfo
  {booktitle} {Pattern Recognition and Tracking XXXI}}},\ \bibinfo {series}
  {Society of Photo-Optical Instrumentation Engineers (SPIE) Conference
  Series}, Vol.\ \bibinfo {volume} {11400}\ (\bibinfo {year} {2020})\ p.\
  \bibinfo {pages} {114000U}\BibitemShut {NoStop}%
\bibitem [{\citenamefont {Chira}\ \emph {et~al.}(2022)\citenamefont {Chira},
  \citenamefont {Haralampiev}, \citenamefont {Winther}, \citenamefont
  {Dittadi},\ and\ \citenamefont
  {Liévin}}]{https://doi.org/10.48550/arxiv.2203.09445}%
  \BibitemOpen
  \bibfield  {author} {\bibinfo {author} {\bibfnamefont {D.}~\bibnamefont
  {Chira}}, \bibinfo {author} {\bibfnamefont {I.}~\bibnamefont {Haralampiev}},
  \bibinfo {author} {\bibfnamefont {O.}~\bibnamefont {Winther}}, \bibinfo
  {author} {\bibfnamefont {A.}~\bibnamefont {Dittadi}}, \ and\ \bibinfo
  {author} {\bibfnamefont {V.}~\bibnamefont {Liévin}},\ }\href {\doibase
  10.48550/ARXIV.2203.09445} {\enquote {\bibinfo {title} {Image
  super-resolution with deep variational autoencoders},}\ } (\bibinfo {year}
  {2022})\BibitemShut {NoStop}%
\end{thebibliography}
%

\end{document}